\documentclass[sigconf]{acmart}

\copyrightyear{2018}
\acmYear{2018}
\setcopyright{acmlicensed}
\acmConference[CCS '18]{2018 ACM SIGSAC Conference on Computer and Communications Security}{October 15--19, 2018}{Toronto, ON, Canada}
\acmBooktitle{2018 ACM SIGSAC Conference on Computer and Communications Security (CCS '18), October 15--19, 2018, Toronto, ON, Canada}
\acmPrice{15.00}
\acmDOI{10.1145/3243734.3243739}
\acmISBN{978-1-4503-5693-0/18/10}

\makeatletter                   
\def\mdseries@tt{m}             
\makeatother                    

\usepackage{booktabs} %
\usepackage{endnotes}
\usepackage{listings}
\usepackage{color}
\usepackage{hyperref}           
\hypersetup{
    colorlinks=true,
    linkcolor=blue,
    filecolor=red,      
    urlcolor=magenta,
    breaklinks=true,            
}
\usepackage{breakurl}           
\usepackage{tabularx}
\usepackage{graphicx}
\usepackage{xspace}
\usepackage{syntax}
\usepackage{amsmath}
\usepackage{adjustbox}
\usepackage{xpatch}
\usepackage{subcaption}
\usepackage{listings}
\usepackage{multirow}
\usepackage{tcolorbox}
\tcbuselibrary{listings,skins}
\usepackage{fixltx2e}
\usepackage{amssymb}
\usepackage{pifont}
\usepackage{tikz}
\usepackage{amsmath}
\usepackage{makecell}%
\usepackage{cases}
\usetikzlibrary{calc,arrows,decorations,patterns,trees,calc,arrows,fit,positioning,shapes,decorations.pathreplacing}
\usepackage[frozencache,cachedir=.]{minted}
\usepackage{algpseudocode}
\usepackage[ruled, vlined, linesnumbered]{algorithm2e}
\newcommand\tikzmark[3][]{%
    \tikz[remember picture,overlay,#1]\node (#2) {#3};%
}

\newcommand{\cmark}{\ding{51}\xspace}
\newcommand{\xmark}{\ding{55}\xspace}

\graphicspath{ {images/} }

\newcommand{\sysname}{BOPC\xspace}
\newcommand{\aref}[1]{\hyperref[#1]{Appendix~\ref*{#1}}}

\newcommand{\ok}{\ding{51}\xspace}
\newcommand{\xA}{\ding{55}\textsubscript{1}\xspace}
\newcommand{\xB}{\ding{55}\textsubscript{2}\xspace}
\newcommand{\xC}{\ding{55}\textsubscript{3}\xspace}
\newcommand{\xD}{\ding{55}\textsubscript{4}\xspace}

\begin{document}
\sloppy
\title{Block Oriented Programming: Automating Data-Only Attacks}
 \author{Kyriakos K. Ispoglou}
 \email{ispo@purdue.edu}
 \affiliation{Purdue University}
 \author{Bader AlBassam}
 \email{balbassa@purdue.edu}
 \affiliation{Purdue University}
 \author{Trent Jaeger}
 \email{tjaeger@cse.psu.edu}
 \affiliation{Pennsylvania State University}
 \author{Mathias Payer}
 \email{mathias.payer@nebelwelt.net}
 \affiliation{EPFL and Purdue University}

\begin{abstract}

With the widespread deployment of Control-Flow Integrity (CFI), control-flow hijacking attacks, and consequently code reuse attacks, are significantly
more difficult.
CFI limits control flow to well-known locations, severely restricting arbitrary
code execution. Assessing the \emph{remaining attack surface} of an application
under advanced control-flow hijack defenses such as CFI and shadow stacks
remains an open problem.

We introduce \sysname, a mechanism to automatically assess 
whether an attacker can execute
arbitrary code on a binary hardened with CFI/shadow stack defenses. \sysname
computes exploits for a target program from payload specifications
written in a Turing-complete, high-level language called SPL that
abstracts away architecture and program-specific details. %
SPL payloads are compiled into a program trace that
executes the desired behavior on top of the target binary. The input
for \sysname is an SPL payload, a starting point (e.g., from a fuzzer
crash) and an arbitrary memory write primitive that allows application
state corruption.  To map SPL payloads to a program trace, \sysname
introduces {\it Block Oriented Programming} (BOP), a new code reuse
technique that utilizes entire basic blocks as gadgets along valid
execution paths in the program, i.e., without violating CFI or shadow
stack policies.
We find that the problem of mapping payloads to program traces is
NP-hard, so \sysname first reduces the search space by pruning
infeasible paths and then uses heuristics to guide the search to
probable paths.  \sysname encodes the BOP payload as a set of memory
writes.

We execute $13$ SPL payloads applied to $10$ popular applications. \sysname
successfully finds payloads and complex execution traces -- which would likely
not have been found through manual analysis -- while following the target's
Control-Flow Graph under an ideal CFI policy in $81\%$ of the cases.
\end{abstract}

\maketitle

\section{Introduction} \label{sec:intro}

Control-flow hijacking and code reuse attacks have been challenging problems for
applications written in C/C++ despite the development and deployment of several
defenses.
Basic mitigations include Data Execution Prevention (DEP)~\cite{DEP} to stop code
injection, Stack Canaries~\cite{GS} to stop stack-based buffer overflows,
and Address Space Layout Randomization (ASLR)~\cite{ASLR} to probabilistically
make code reuse attacks harder. These mitigations can be bypassed through, e.g.,
information leaks~\cite{aslrbypass, aslrbypass2, gsbypass, gsbypass2} or code
reuse attacks~\cite{advr2l, ROP, JOP, shacham, katochwhitepaper}.

Advanced control-flow hijacking defenses such as Control-Flow Integrity
(CFI)~\cite{CFI, LLVM-CFI, MSCFI, burow17csur} or shadow stacks/safe
stacks~\cite{shadow, CPI} limit the set of allowed target addresses for
indirect control-flow transfers. CFI mechanisms typically rely on static
analysis to recover the Control-Flow Graph (CFG) of the application.  These
analyses over-approximate the allowed targets for each indirect dispatch
location. At runtime, CFI checks determine if the observed target for each
indirect dispatch location is within the allowed target set for that 
dispatch location as identified by the CFG analysis.
Modern CFI
mechanisms~\cite{LLVM-CFI, MSCFI, MCFI, PICFI} are deployed in, e.g., Google
Chrome~\cite{cfichrome}, Microsoft Windows 10, and Edge~\cite{cfguard}.

However, CFI still allows the attacker control over the execution along two
dimensions: first, due to imprecision in the analysis and CFI's statelessness, 
the attacker can choose any of the targets in the set for each dispatch; second,
data-only attacks allow an attacker to influence conditional branches
arbitrarily.
Existing attacks against CFI leverage manual analysis to construct exploits for
specific applications along these two dimensions~\cite{cfb, COOP, jujutsu, ooc,
stitching}. With CFI, exploits become highly program dependent as the set of
reachable gadgets is severely limited by the CFI policy,
so exploits must therefore follow valid paths in the CFG. Finding a path along
the CFG that achieves the exploit goals is much more complex than simply finding
the locations of gadgets. As a result, building attacks against advanced
control-flow hijacking defenses has become a challenging, predominantly manual
process.

We present \sysname (\emph{Block Oriented Programming Compiler}) , 
an automatic framework to evaluate a program's remaining
attack surface under strong control-flow hijacking mitigations. \sysname
automates the task of finding an execution trace through a buggy program that
executes arbitrary, attacker-specified behavior.
\sysname compiles an ``exploit'' into a program trace, which is executed on
top of the original program's CFG. To express the
desired exploits flexibly, \sysname provides a Turing-complete,
high-level language: SPloit Language (SPL).  To interact with the
environment, SPL provides a rich API to call OS functions, direct
access to memory, and an abstraction for hardware registers.
\sysname takes as input an SPL payload and
a starting point (e.g., found through fuzzing or manual analysis) and
returns a trace through the program (encoded as a set of memory
writes) that encodes the SPL payload.

The core component of \sysname is the mapping process through a novel code reuse
technique we call \emph{Block Oriented Programming} (BOP). First, \sysname
translates the SPL payload into constraints for individual statements and, for
each statement, searches for basic blocks in the target binary that satisfy
these constraints (called \emph{candidate blocks}). At this point, SPL abstracts register
assignments from the underlying architecture. Second, \sysname infers a
resource (register and state) mapping for each SPL statement, iterating through
the set of candidate blocks and turning them into \emph{functional blocks}.
Functional blocks can be used to execute a concrete instantiation of the given
SPL statement.  Third, \sysname constructs a trace that connects each functional
block through \emph{dispatcher blocks}. Since the mapping process is NP-hard, to find
a solution in reasonable time \sysname first prunes the set of functional blocks
per statement to constrain the search space and then uses a ranking based on the
proximity of individual functional blocks as a heuristic when searching for
dispatcher gadgets.

We evaluate \sysname on $10$ popular network daemons and setuid programs,
demonstrating that \sysname can generate traces from a set of $13$ test
payloads.  Our test payloads are both reasonable exploit payloads (e.g., calling
\texttt{execve} with attacker-controlled parameters) as well as a demonstration
of the computational capabilities of SPL (e.g., loops and conditionals).
Applications of \sysname go beyond an attack framework.  We envision \sysname as
a tool for defenders and software developers to highlight the \emph{residual}
attack surface of a program.  %
For example, a developer can test whether a bug at a particular statement
enables a practical code reuse attack in the program.  Overall, we present the
following contributions:

\begin{itemize}
  \item \emph{Abstraction:} We introduce SPL, a C dialect with
access to virtual registers and an API to call OS and other
library functions, suitable for writing exploit payloads. SPL enables the
necessary abstraction to scale to large applications.

  \item \emph{Search:} Development of a \emph{trace module} that allows
execution of an arbitrary payload, written in SPL, using the target binary's
code. The trace module considers strong defenses such as DEP, ASLR, shadow
stacks, and CFI alone or in combination. The trace module enables the
discovery of viable mappings through a search process.

  \item \emph{Evaluation:} Evaluation of our prototype demonstrates the generality of
our mechanism and uncovers exploitable vulnerabilities where manual exploitation
may have been infeasible.  For $10$ target programs, \sysname successfully
generates exploit payloads and program traces to implement code reuse attacks
for $13$ SPL exploit payloads for 81\% of the cases.

\end{itemize}

\section{Background and Related Work} \label{sec:BG}

Initially, exploits relied on simple code injection to execute arbitrary code.
The deployment of Data Execution Prevention (DEP)~\cite{DEP} mitigated code
injection and attacks moved to {\it reusing} existing code.  The first code
reuse technique, {\it return to libc}~\cite{designer1997return}, simply reused
existing libc functions.  {\it Return Oriented Programming} (ROP)~\cite{ROP}
extended code reuse to a Turing-complete technique.  ROP locates small sequences
of code which end with a return instruction, called ``gadgets.'' Gadgets are
connected by injecting the correct state, e.g., by preparing a set of invocation
frames on the stack~\cite{ROP}. A number of code reuse variations
followed~\cite{JOP, noretrop, homescu2012microgadgets}, extending the approach
from return instructions to arbitrary indirect control-flow transfers.

Several tools~\cite{ropgadget, q, PSHAPE, ropc} seek to automate ROP payload
generation. However, the automation suffers from inherent limitations.
These tools fail to find gadgets in the target binary that do not follow the
expected form ``\texttt{inst1; inst2; ... retn;}'' as they search for a
set of hard coded gadgets that form pre-determined gadget chains. Instead of
abstracting the required computation, they search for specific gadgets.
If any
gadget is not found or if a more complex gadget chain is needed, these tools
degenerate to gadget dump tools, leaving the process of gadget chaining to the
researcher who manually creates exploits from discovered gadgets.

The invention of code reuse attacks resulted in a plethora of new detection
mechanisms based on execution anomalies and heuristics~\cite{ropdetect1,
ropdetect2, ropdetect3, ropdetect4, ropdetect5} such as frequency of return
instructions. Such heuristics can often be bypassed~\cite{ropattack}. 

While the aforementioned tools help to craft appropriate payloads, finding the
vulnerability is an orthogonal process.  Automatic Exploit Generation
(AEG)~\cite{AEG} was the first attempt to automatically find vulnerabilities and
generate exploits for them. AEG is limited in that it does not assume any
defenses (such as the now basic DEP or ASLR mitigations).  The generated
exploits are therefore buffer overflows followed by static shellcode.

\subsection{Control Flow Integrity}
Control Flow Integrity~\cite{CFI, LLVM-CFI, MSCFI, burow17csur} (CFI) {\it
mitigates} control-flow hijacking to arbitrary locations (and therefore 
code reuse attacks). CFI restricts the set of potential targets that are 
reachable from an indirect dispatch. While CFI does not stop the initial memory
corruption, it validates the code pointer before it is used. CFI infers an
(overapproixmate)
CFG of the program to determine the allowed targets for
each indirect control-flow transfer.  Before each indirect dispatch, the target
address is checked to determine if it is a valid edge in the CFG, and if not an
exception is thrown. This limits the freedom for the attacker, as she can only
target a small set of targets instead of any executable byte in memory.
For example, an attacker may overwrite a function pointer through a buffer
overflow, but the function pointer is checked before it is used. Note that CFI
targets \emph{forward edges}, i.e., virtual dispatchers for C++ or indirect
function calls for C.

With CFI, code reuse attacks become harder, but not impossible~\cite{cfb,
jujutsu, ooc, COOP}.  Depending on the application and strength of the CFI
mechanism, CFI can be bypassed with Turing-complete payloads, which are often 
highly complex to comply with the CFG. So far, these code-reuse
attacks rely on manually constructed payloads.

Deployed CFI implementations~\cite{FGCFI, LLVM-CFI,MCFI, MSCFI, PICFI} use a
static over-approximation of the CFG based on method prototypes and class
hierarchy. PittyPat~\cite{PittyPat} and PathArmor~\cite{PathArmor} introduce
path sensitivity that evaluates partial execution paths. Newton~\cite{Newton}
introduced a framework that reasons about the strength of defenses, including
CFI. Newton exposes indirect pointers (along with their allowed target set) that
are reachable (i.e., controllable by an adversary) through given entry points.
While Newton displays all usable ``gadgets,'' it cannot stitch them together and
effectively is a CFI-aware ROP gadget search tool that helps an analyst to
manually construct an attack.

\subsection{Shadow Stacks}

While CFI protects {\it forward} edges in the CFG (i.e., function pointers or
virtual dispatch), a shadow stack orthogonally protects {\it backward} edges
(i.e., return addresses). Shadow stacks keep a protected copy
(called \emph{shadow}) of all return addresses on a separate, protected stack.
Function calls store the return address both on the regular stack and on the
shadow stack. When returning from a function, the mitigation checks for
equivalence and reports an error if the two return addresses do not match.
The shadow stack itself is assumed to be at a protected memory location to keep
the adversary from tampering with it.
Shadow stacks enforce stack integrity and protect the binary from any
control-flow hijacking attack against the backward edge.

\subsection{Data-only Attacks}

While CFI mitigates code-reuse attacks, CFI cannot stop data-only attacks.
Manipulating a program's \emph{data} can be enough for a successful
exploitation. Data-only attacks target the program's data rather than its
control flow. E.g., having full control over the arguments to
\texttt{execve()} suffices for arbitrary command execution. Also, data in a
program may be sensitive: consider overwriting the \texttt{uid} or a variable
like \texttt{is\_admin}. {\it Data Oriented Programming} (DOP)~\cite{dop} is the
generalization of data-only attacks. Existing DOP attacks rely on an analyst to
identify sensitive variables for manual construction.

Similarly to CFI, it is possible to build the {\it Data Flow Graph} of the program
and apply {\it Data Flow Integrity} (DFI)~\cite{DFI} to it. However, to the
best of our knowledge, there are no practical DFI-based defenses due to
prohibitively high overhead of data-flow tracking.

In comparison to existing data-only attacks, \sysname automatically generates
payloads based on a high-level language. The payloads follow the
valid CFG of the program but not its Data Flow Graph.

\section{Assumptions and Threat Model} \label{sec:threat}

Our threat model consists of a binary with a known memory corruption
vulnerability that is protected with the state-of-the-art control-flow hijack
mitigations, such as CFI along with a Shadow Stack.
Furthermore, the binary is also hardened with DEP, ASLR and Stack Canaries.	

We assume that the target binary has an arbitrary memory write vulnerability.
That is, the attacker can write \emph{any} value to \emph{any} (writable) 
address.  We call this an {\it Arbitrary memory Write Primitive} (AWP).
To bypass probabilistic defenses such as ASLR, %
we assume that the attacker
has access to an information leak, i.e., a vulnerability that allows her to
read \emph{any} value from \emph{any}  memory address. We call this an {\it
Arbitrary memory Read Primitive} (ARP). Note that the ARP is optional and
only needed to bypass orthogonal probabilistic defenses.

We also assume that there exists an entry point, i.e., a location that the
program reaches naturally after completion of all AWPs (and ARPs). Thus \sysname
does \emph{not} require code pointer corruption to reach the entry point.
Determining an entry point is considered to be part of the vulnerability
discovery process. Thus, finding this entry point is orthogonal to our work.

Note that these assumptions are in line with the threat model of control-flow
hijack mitigations that aim to prevent attackers from exploiting arbitrary read
and write capabilities.
These assumptions are also practical. Orthogonal bug finding tools such as
fuzzing often discover arbitrary memory accesses that can be abstracted to the
required arbitrary read and writes, placing the entry point right after the AWP.
Furthermore, these assumptions map to real bugs.
Web servers, such as nginx, spawn threads to handle requests and a bug in the
request handler can be used to read or write an arbitrary memory address. Due
to the request-based nature, the adversary can repeat this process multiple
times. After the completion of the state injection, the program follows an
alternate and disjoint path to trigger the injected payload.

These assumptions enable \sysname to inject a payload into a target binary's
address space, modifying its memory state to execute the payload.
\sysname assumes that the AWP (and/or
ARP) may be triggered multiple times to modify the memory state of
the target binary. After the state modification completes, the SPL payload
executes without using the AWP (and/or ARP) further. This separates
SPL execution into two phases: state modification and payload execution. The
AWP allows state modification, \sysname infers the required state
change to execute the SPL payload.

\section{Design} \label{sec:design}

\begin{figure}[!b]
  \centering
\vspace{-1em}
    \begin{tikzpicture}
        [
            line width=1pt, >=stealth',
            tall/.style={align=center,minimum height=80pt},
            mylabel/.style={midway,fill=white,inner sep=1pt},
            font=\sffamily,
            color=gray
        ]

        \tikzset{main node/.style={
            align=center,
            rectangle,draw,color=black, minimum height=25pt, minimum width=80pt,
        },label node/.style={
            draw=none,
            font=\bfseries
        }}

        \tikzstyle{myarrow}=[->, >=latex, thick,color=black]

        \node[main node] (Payloads) [] {(1) SPL Payload};
        \node[main node] (Mappings) [right=of Payloads] {(2) Selecting\\functional blocks};
        \node[main node] (FindDispatcher) [below=10pt of Mappings] {(3) Searching for\\dispatcher blocks};
        \node[main node] (ComputePayload) [left=of FindDispatcher] {(4) Stitching\\BOP gadgets};

        \draw[myarrow] (Payloads) -- (Mappings);
        \draw[myarrow] (Mappings) -- (FindDispatcher);
        \draw[myarrow] (FindDispatcher) -- (ComputePayload);

    \end{tikzpicture}
  \caption{Overview of \sysname's design.
  }
  \label{fig:doverview}
\end{figure}
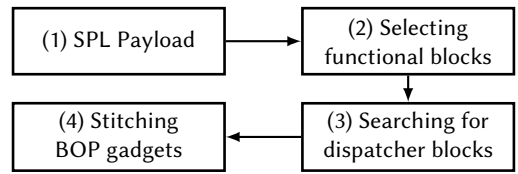

\autoref{fig:doverview} shows how \sysname automates the analysis tasks
necessary to leverage AWPs to produce a useful exploit in the presence of strong
defenses, including CFI.  First, \sysname provides an exploit programming
language, called SPL, that enables analysts to define
exploits independent of the target program or underlying architecture.  Second,
to automate SPL gadget discovery, \sysname finds basic blocks from the target
program that implement individual SPL statements, called {\em functional
blocks}.  Third, to chain basic blocks together in a manner that adheres with
CFI and shadow stacks, \sysname searches the target program for sequences of
basic blocks that connect pairs of neighboring functional blocks, which we call
{\em dispatcher blocks}.  Fourth, \sysname simulates the BOP chain to produce a
payload that implements that SPL payload from a chosen AWP.

\begin{figure}[t]
    \centering
\resizebox{0.78\linewidth}{!} {

\begin{tikzpicture}
    [
        line width=0.5pt, >=stealth', level distance=0.35cm,
        functional/.style={pattern=north east lines,pattern color=gray,rectangle,draw,align=center,minimum height=10pt,minimum width=40pt,thick,anchor=north},
        dispatcher/.style={fill=none,thick,rectangle,draw,minimum width=20pt, anchor=north},
        negated/.style={fill=none,cross out,draw, anchor=north,scale=0.5},
        edge from parent path={(\tikzparentnode.south) -- ++(0,-.07cm)
        -| (\tikzchildnode.north)},
    ]

    \tikzstyle{myarrow}=[->, >=open triangle 90, thick]

    \node[functional] (root) {}
    child{
        node[dispatcher]{}
        child{
            node[dispatcher] {}{
                child{
                    node[negated] {}
                }
                child{
                    node[dispatcher] (tail) {}
                    child{node[dispatcher] {}
                    child{node[dispatcher] {}
                    child{node[functional] (newroot) {}}
                }
                    child{node[negated] {}
                }
            }
            child{
                node[negated] {}}
            }
        }}
        child{node[negated] {}
}
}
child{node[negated] {}
    }
    ;

    \node[fit=(root.north east)(newroot.north west)](box){};
    \draw[thick] ([xshift=5pt,yshift=3pt]root.north east) rectangle ([xshift=-5pt,yshift=3pt]newroot.north west);
    \node (legend1) [left=58pt of root, rectangle, draw, pattern=north east lines, pattern color=gray,scale=0.7]{};
    \node (legend2) [below=3pt of legend1, rectangle, draw, scale=0.7]{};
    \node (description1) [right=0pt of legend1, scale=0.5] {Functional};
    \node (description2) [right=0pt of legend2, scale=0.5] {Dispatcher};

    \node[fit=(legend1)(legend2)(description1)(description2)](legend){};
    \draw[thin]([yshift=-3pt,xshift=-2pt]legend.north east) rectangle ([yshift=3pt,xshift=2pt]legend.south west);

    \draw [decorate,decoration={brace,amplitude=10pt,mirror,raise=4pt},yshift=0pt]
    ([yshift=6pt]box.south east) -- ([yshift=-0.5pt]box.north east) node [black,midway,xshift=1.0cm,align=left,font=\footnotesize] {BOP\\Gadget};

\end{tikzpicture}
}
\caption{BOP gadget structure. The {\it functional} part consists of a single 
basic block that executes an SPL statement. Two functional blocks are chained
together through a series of {\it dispatcher} blocks, without clobbering
the execution of the previous functional blocks.} \label{fig:gadget}
\end{figure}
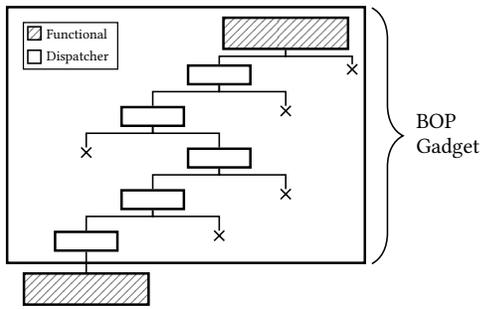

The \sysname design builds on two key ideas: {\em 
Block Oriented Programming} and
{\em Block Constraint Summaries}.  First, defenses such as CFI impose stringent
restrictions on transitions between gadgets, so an exploit no longer has the
flexibility of setting the instruction pointer to arbitrary values.  Instead,
\sysname implements {\it Block Oriented Programming} (BOP), which constructs
exploit programs called {\em BOP chains} from basic block sequences in the valid
CFG of a target program.  Note that our CFG encodes both {\it forward edges}
(protected by CFI) and {\it backward edges} (protected by shadow stack).
For BOP, gadgets are chains of {\it entire} basic blocks (sequences of
instructions that end with a direct \emph{or} indirect control-flow transfer),
as shown in \autoref{fig:gadget}. A BOP chain consists of a sequence of {\em BOP
gadgets} where each BOP gadget is: one \emph{functional block} that implements a
statement in an SPL payload and zero or more \emph{dispatcher blocks} that
connect the functional block to the next BOP gadget in a manner that complies
with the CFG.

Second, \sysname abstracts each basic block from individual instructions into
{\em Block Constraint Summaries}, enabling blocks to be employed in a variety of
different ways.  That is, a single block may perform multiple functional and/or
dispatching operations by utilizing different sets of registers for different
operations.
That is, a basic block that modifies a register in a manner that may fulfill an
SPL statement may be used as a functional block, otherwise it may be considered
to serve as a dispatcher block. 

\sysname leverages abstract Block Constraint Summaries to apply blocks in
multiple contexts. At each stage in the development of a BOP chain, the blocks
that may be employed next in the CFG as dispatcher blocks to connect two
functional blocks depend on the block summary constraints for each block.
There are two cases: either the candidate dispatcher block's summary 
constraints
indicate that it will modify the register state set and/or the memory state 
by the functional blocks,
called the {\em SPL state}, or it will not, enabling the computation to proceed
without disturbing the effects of the functional blocks. A block that modifies
a current SPL state unintentionally, is said to be a {\em clobbering block} 
for that state.
Block summary constraints enable identification of clobbering blocks at each
point in the search.

An important distinction between BOP and conventional ROP (and variants) is that
the problem of computing BOP chains is NP-hard, as proven in
\autoref{app:np-proof}. Conventional ROP assumes that indirect control-flows may
target any executable byte in memory while BOP must follow a legal path through
the CFG for any chain of blocks, resulting in the need for automation.

\lstset{
 language=C,
 morekeywords={returnto, int64, string}
}

\begin{table}[t]
\centering
\begin{adjustbox}{width=0.9\linewidth}
\begin{tabular}{l|l}
\multicolumn{1}{c|}{\it Simple loop} &
\multicolumn{1}{c}{\it Spawn a shell} \\ \hline
\begin{lstlisting}
void payload() {
  __r0 = 0;
  
LOOP:
  __r0 += 1;
  if (__r0 != 128)
    goto LOOP;
      
  returnto 0x446730;
}
\end{lstlisting}      
&
\begin{lstlisting}
void payload() {
  string prog = "/bin/sh\0";
  int64 *argv = {&prog, 0x0};    
  
  __r0 = &prog;
  __r1 = &argv;
  __r2 = 0;

  execve(__r0, __r1, __r2);
}
\end{lstlisting}
\\ %
\end{tabular}
\end{adjustbox}
\caption{Examples of SPL payloads.} \label{tab:examples}
\vspace{-1em}
\end{table}

\subsection{Expressing Payloads} \label{sec:spl}

\sysname provides a programming language, called {\it SPloit Language} (SPL)
that allows analysts to express exploit payloads in a compact high-level
language that is independent of target programs or processor architectures.  SPL
is a dialect of C. Compared to minDOP~\cite{dop}, SPL allows use of both virtual
registers and memory for operations and declaration of variables/constants.
\autoref{tab:examples} shows some sample payloads.  Overall, SPL has the
following features:

\begin{itemize}
  \item It is Turing-complete;
  \item It is architecture independent;
  \item It is close to a well known, high level language.
\end{itemize}

Compared to existing exploit development tools~\cite{q, PSHAPE, ropgadget}, the
architecture independence of SPL has important advantages. First, the same
payload can be executed under different ISAs or operating systems.
Second, SPL uses a set of \emph{virtual registers}, accessed through reserved
volatile variables. Virtual registers increase flexibility, which in turn
increases the chances of finding a solution: virtual registers may be mapped to
any general purpose register and the mapping may be changed dynamically.

To interact with the environment, SPL defines a concise API to access OS
functionality.  Finally, SPL supports conditional and unconditional jumps to
enable control-flow transfers to arbitrary locations. This feature makes SPL a
Turing-complete language, as proven in \aref{app:spl-proof}. The complete
language specifications are shown in \aref{app:ebnf} in Extended Backus--Naur
form (EBNF).

The environment for SPL differs from that of conventional languages. Instead of
running code directly on a CPU, our compiler encodes the payload as a mapping of
instructions to functional blocks.  That is, the underlying runtime environment
is the target binary and its program state, where payloads are executed as side
effects of the underlying binary.

\vspace{-0.1in}
\subsection{Selecting functional blocks} \label{sec:functional}
To generate a BOP chain for an SPL payload, \sysname must find a
sequence of blocks that implement each statement in the SPL payload,
which we call {\em functional blocks}.  The process of
building BOP chains starts by identifying functional blocks per
SPL statement.

Conceptually, \sysname must compare each block to each SPL statement
to determine if the block can implement the statement.  However,
blocks are in terms of machine code and SPL statements are
high-level program statements.  To provide flexibility for matching
blocks to SPL statements, \sysname computes {\em Block Constraint
  Summaries}, which define the possible impacts that the block would
have on SPL state.  Block Constraint Summaries provide flexibility in
matching blocks to SPL statements because there are multiple possible
mappings of SPL statements and their virtual registers to the block
and its constraints on registers and state.

The constraint summaries of each basic block are obtained by {\it
  isolating} and {\it symbolically} executing it.  The effect of
symbolically executing a basic block creates a set of constraints, mapping
input to the resultant output.  Such constraints refer to
registers, memory locations, jump types and external operations (e.g., library
calls).

To find a match between a block and an SPL statement the block must
perform all the operations required for that SPL statement. More
specifically, the constraints of the basic block should contain all
the operations required to implement the SPL statement.

\subsection{Finding BOP gadgets} \label{sec:find}

\sysname computes a {\it set} of all \emph{potential} functional
blocks for each SPL statement or halts if any statement has no blocks.
To stitch functional blocks, \sysname must select one
functional block and a sequence of dispatcher blocks that reach the
next functional block in the payload.  The combination of a functional
block and its dispatcher blocks is called a {\em BOP gadget}, as
shown in \autoref{fig:gadget}. To build a BOP gadget, \sysname
must select {\it exactly} one functional block from each set and find
the appropriate dispatcher blocks to connect to a subsequent functional block.

However, dispatcher paths between two functional blocks may not exist
either because there is no legal path in the CFG between them,
or the control flow cannot reach the next block
due to unsatisfiable runtime constraints.
This constraint imposes limits on functional block selection, as
the existence of a dispatcher path depends on the \emph{previous} BOP gadgets.

\begin{figure}[t]
    \begin{subfigure}[b]{0.28\linewidth}
    \end{subfigure}
    \begin{subfigure}[b]{0.28\linewidth}
        \includegraphics[width=\textwidth]{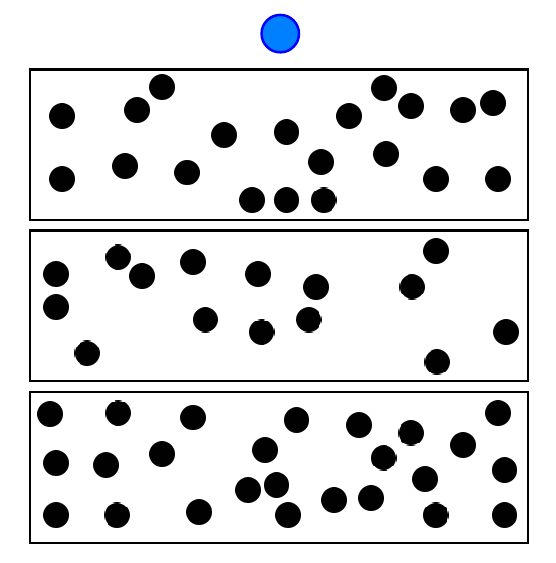}
        \caption{}
        \label{fig:stitch01}
    \end{subfigure}
    \hfill%
    \begin{subfigure}[b]{0.28\linewidth}
        \includegraphics[width=\textwidth]{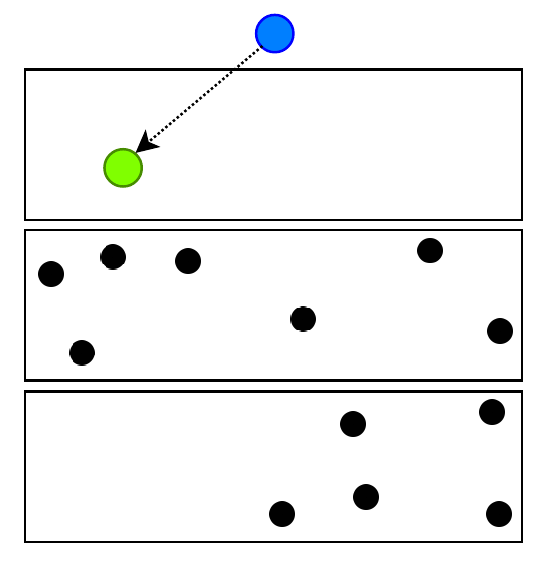}
        \caption{}
        \label{fig:stitch02}
    \end{subfigure}
    \hfill%
    \begin{subfigure}[b]{0.28\linewidth}
        \includegraphics[width=\textwidth]{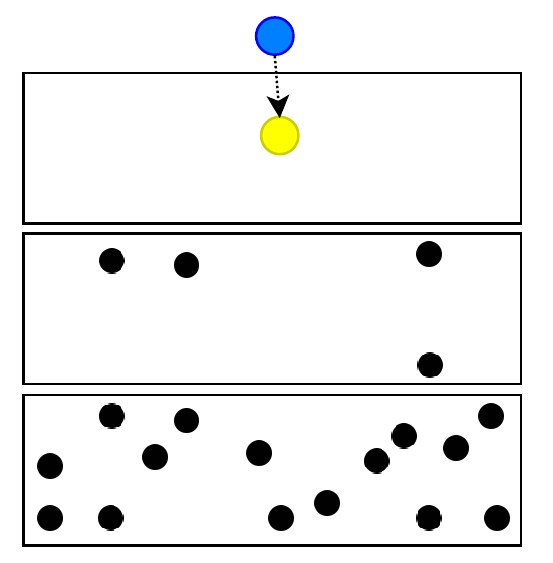}
        \caption{}
        \label{fig:stitch03}
    \end{subfigure}
    \caption{Visualisation of BOP gadget volatility, rectangles: SPL
      statements, dots: functional blocks (a).
      Connecting any two statements through dispatcher
      blocks constrains remaining gadgets (b), (c).
    	}\label{fig:bopGadgetStitching}
\vspace{-1em}
\end{figure}

BOP gadgets are \emph{volatile}: gadget feasibility changes based on
the selection of prior gadgets for the target binary. This is
illustrated in \autoref{fig:bopGadgetStitching}.
The problem of selecting a suitable sequence of functional blocks,
such that a dispatcher path exists between every possible control flow
transfer in the SPL payload, is NP-hard, as we prove in
\aref{app:np-proof}. Even worse, an approximation algorithm does not
exist.

As the problem is unsolvable in polynomial time in the general case, we propose
several heuristics and optimizations to find solutions in reasonable amounts of
time. \sysname leverages basic block {\it proximity} as a metric to ``rank''
dispatcher paths and organizes this information into a special data structure,
called a \emph{delta graph} that provides an efficient way to probe potential
sequences of functional blocks.

\begin{figure}[t]
\begin{lstlisting}[escapeinside={(*@}{@*)}]
    (*@\textbf{Function\_1:}@*)
      <instructions>
      ...
      call Function_2        (*@\textbf{Function\_2:}@*)
      (*@\tikzmark{InsAfterFCBarLeft}{}@*)<insn_after_call>(*@\tikzmark{InsAfterFCBarRight}{}@*)        (*@\tikzmark{BarLeft}{}@*)<prologue>
      ...                      ...
    (*@\tikzmark{BstartLeft}{}@*)B:
      <instructions>           <instructions>
    (*@\tikzmark{AstartLeft}{}@*)A:
      <nop_sled>               ...
      (*@\tikzmark{ACBarLeft}{}@*)call Function_2(*@\tikzmark{ACBarRight}{}@*)          (*@\tikzmark{RetLeft}{}@*)retn(*@\tikzmark{RetRight}{}@*)
      (*@\tikzmark{InsAfterACBarLeft}{}@*)<insn_after_call>(*@\tikzmark{InsAfterACBarRight}{}@*)
      retn

(*@
\begin{tikzpicture}[remember picture, overlay]
    \draw[->, very thick, red, out=180, in=150, looseness=2] (InsAfterFCBarLeft) to (BstartLeft);
    \draw[->, very thick, red, out=180, in=180, looseness=2] (AstartLeft) to (ACBarLeft);
    \draw[->, very thick, red, out=180, in=0] (RetLeft) to (InsAfterFCBarRight);
    \draw[->, very thick, red, out=0,   in=180] (ACBarRight) to (BarLeft);
    \draw[->, dashed, ultra thick, blue, out=180, in=0] (RetLeft.west) to (InsAfterACBarRight);

    \node[draw,red,circle,thick,inner sep=1.5pt] (T1) at ([xshift=-10pt,yshift=-25pt]$(AstartLeft)!0.5!(AstartLeft)$) {1};
    \node[draw,red,circle,thick,inner sep=1.5pt] (T4) at ([xshift=-10pt,yshift=25pt]$(BstartLeft)!0.5!(BstartLeft)$) {4};
    \node[draw,red,circle,thick,inner sep=1.5pt] (T2) at ([xshift=12pt,yshift=23pt]$(ACBarRight)!0.5!(ACBarRight)$) {2};
    \node[draw,red,circle,thick,inner sep=1.5pt] (T3) at ([xshift=-8pt,yshift=23pt]$(RetLeft)!0.5!(RetLeft)$) {3};
\end{tikzpicture}
@*)
    \end{lstlisting}
    \vspace{-22pt}
\caption{Existing shortest path algorithms are unfit to 
   	measure proximity in the CFG. Consider the shortest path from \texttt{A}
   	to \texttt{B}. A context-unaware shortest path algorithm will mark the red
path as solution,
instead of following the blue arrow upon return from 
    \texttt{Function_2}, it follows the red arrow (3).}
 \vspace{-0.15in}
\label{fig:cfgdist}	
\end{figure}
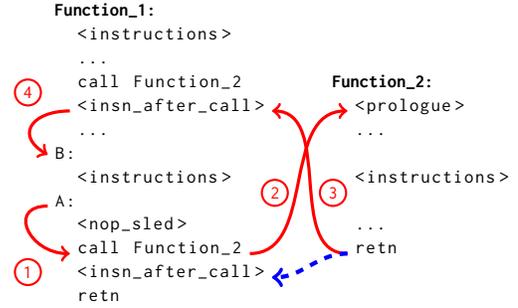

\subsection{Searching for dispatcher blocks} \label{sec:delta}

While each functional block executes a statement, \sysname must chain
multiple functional blocks together to execute the SPL payload.
Functional blocks are connected through zero or more basic blocks that do not
clobber the SPL state computed thus far. Finding such non-clobbering
blocks that transfer control from one functional statement to another
is challenging as each additional block increases the constraints and
path dependencies.  
Thus, we propose a graph data structure, called the \emph{delta
  graph}, to represent the state of the search for dispatcher blocks.
The delta graph stores, for each functional block for each SPL
statement, the shortest path to the next candidate block.  Stitching
arbitrary sequences of statements is NP-hard as each selected path
between two functional statements influences the availability of
further candidate blocks or paths, we therefore leverage the delta
graph to try \emph{likely} candidates first.

The intuition behind the \emph{proximity} of functional blocks is that shorter
paths result in simpler and more likely satisfiable constraints.  Although
this metric is a \emph{heuristic},
our evaluation (\autoref{sec:eval}) shows that it works well in practice.

The delta graph enables quick elimination of sets of functional blocks that are
highly unlikely to have dispatcher blocks and thus constitute a BOP gadget. For
instance, if there is no valid path in the CFG between two functional blocks
(e.g., if execution has to traverse the CFG ``backwards''), no dispatcher
will exist and therefore, these two functional blocks cannot be part of the
solution.

The delta graph is a multi-partite, directed graph that has a set of
functional block nodes for {\it every} payload statement. 
An edge between two functional blocks represents the \emph{minimum} number
of executed basic blocks to move from one functional block
to the other, while {\it avoiding} clobbering blocks. See
\autoref{fig:deltagraph} for an example.

Indirect control-flow transfers pose an interesting challenge when calculating
the shortest path between two basic blocks in a CFG: while they statically allow
multiple targets, at runtime they are context sensitive and only have one
{\it concrete} target.

Our context-sensitive shortest path algorithm is a {\it recursive}
version of Dijkstra's~\cite{algo} shortest path algorithm that {\it avoids} all
clobbering blocks.. 
Initially, each edge on the CFG has a cost of $1$. 
When it encounters a basic block with a call instruction, it recursively
calculates the shortest paths starting from the calling function's entry block,
$B_E$ (a {\it call stack} prevents deadlocks for recursive callees).
If the destination block, $B_D$, is inside the callee, the shortest path is 
the concatenation of the two individual shortest paths from the beginning to 
$B_E$ and from $B_E$ to $B_D$.
Otherwise, our algorithm finds the \emph{shortest path} from the $B_E$ to the 
closest return point and uses this value as an edge weight for that
callee.

\begin{table}[t]
\centering
\begin{adjustbox}{width=1\linewidth}
\begin{tabular}{l|l}
\multicolumn{1}{c|}{\it Long path with simple constraints} &
\multicolumn{1}{c} {\it Short path with complex constraints} \\ \hline
\begin{lstlisting}[language=C]
a, b, c, d, e = input();
// point A
if (a == 1) {
  if (b == 2) {
    if (c == 3) {
      if (d == 4) {
        if (e == 5) {
          // point B
\end{lstlisting}
&
\begin{lstlisting}[language=C]
a = input();

X = sqrt(a);
Y = log(a*a*a - a)

// point A
if (X == Y) {
  // point B

\end{lstlisting}
\end{tabular}
\end{adjustbox}
\caption{A counterexample that demonstrates why proximity between two 
functional blocks can be inaccurate. Left, we can move 
from point A to point B even if they are 5 blocks apart from each other. 
Right, it is much harder to satisfy the constrains 
and to move from A to B, despite the fact that A and B are only 1 block apart.
\vspace{-2em}}
\label{tab:heuristic} 
\end{table}

After creation of the delta graph, our algorithm selects {\it exactly}
one node (i.e., functional block) from each set (i.e., payload
statement), to \emph{minimize} the total weight of the resulting {\it induced
  subgraph~\footnote{The {\it induced subgraph} of the delta graph is
    a subgraph of the delta graph with one node (functional block) for
    each SPL statement and with edges that represent their shortest
    available dispatcher block chain.}}.
This selection of functional blocks is considered to be the most likely to
give a solution, so the next step is to find the exact dispatcher blocks and
create the BOP gadgets for the SPL payload.

\subsection{Stitching BOP gadgets} \label{sec:stitch}

The minimum induced subgraph from the previous step determines a set of
functional blocks that may be stitched together into an SPL payload. This set of 
functional blocks has minimal distance to each other, thus making satisfiable
dispatcher paths more likely.

To find a dispatcher path between two functional blocks, \sysname leverages
\emph{concolic execution}~\cite{concolic} (symbolic execution along a given
path). Along the way, it collects the required constraints that are needed to
lead the execution to the next functional block.  Symbolic execution
engines~\cite{klee, angr} translate basic blocks into sets of constraints and
use Satisfiability Modulo Theories (SMT) to find satisfying assignments for
these constraints; symbolic execution is therefore NP-complete.  Starting from
the (context sensitive) shortest path between the functional blocks, \sysname
\xspace {\it guides} the symbolic execution engine, collecting the corresponding
constraints. 

To construct an SPL payload from a BOP chain, \sysname launches concolic
execution from the first functional block in the BOP chain, starting with an
empty state. At each step \sysname tries the first $K$ shortest dispatcher paths
until it finds one that reaches the next functional block (the edges in the
minimum induced subgraph indicate which is the ``next'' functional block). The
corresponding constraints are added to the current state.  The search therefore
{\it incrementally} adds BOP gadgets to the BOP chain.
When a functional block represents a conditional SPL statement, its node in the
induced subgraph contains two outgoing edges (i.e., the execution can transfer
control to two different statements). However during the concolic execution, the
algorithm does not know which one will be followed, it {\it clones} the current
state and independently follows both branches, exactly like symbolic
execution~\cite{klee}.

Reaching the last functional block, \sysname checks whether the constraints 
have a satisfying assignment and forms an exploit payload. Otherwise, it falls
back and tries the next possible set of functional blocks.
To repeat that execution on top of the target binary, these constraints 
are concretized and translated into a memory layout that will be initialized 
through AWP in the target binary.

\section{Implementation}

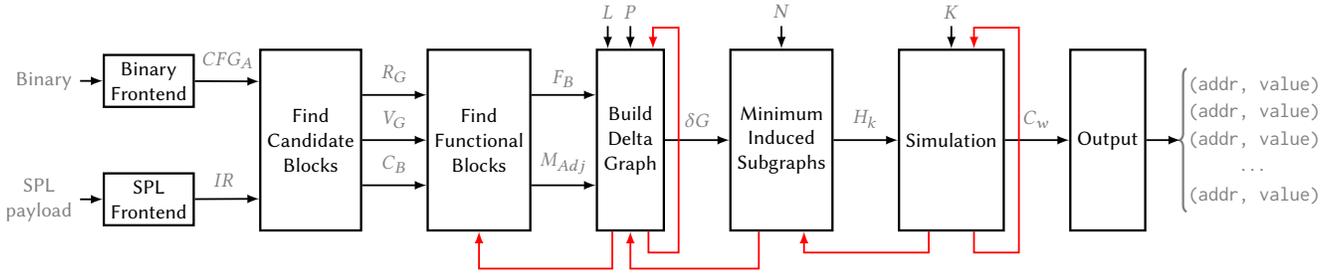
\begin{figure*}[!t]
  \centering
    \resizebox{\textwidth}{!} {
    \begin{tikzpicture}
        [
            line width=1pt, >=stealth',
            tall/.style={align=center,minimum height=80pt},
            mylabel/.style={midway,fill=white,inner sep=1pt},
            font=\sffamily,
            color=gray
        ]

        \tikzset{main node/.style={
            align=center,
            rectangle,draw,color=black
        },label node/.style={
            draw=none,
            font=\bfseries
        }}

        \tikzstyle{myarrow}=[->, >=latex, thick,color=black]

        \node[main node] (BIN-FE) [] {Binary\\Frontend};
        \node[] (BIN-E) [left=10pt of BIN-FE]{Binary};

        \node[main node] (SPL-FE) [below=of BIN-FE] {SPL\\Frontend};
        \node[align=center] (SPL-E) [left=10pt of SPL-FE] {SPL\\payload};

        \node[main node, tall] (FIND-CAND)
        [right= of $(SPL-FE.north east)!0.5!(BIN-FE.south east)$] {Find\\Candidate\\Blocks};

        \node[main node, tall] (FindFuncBlocks) [right=of FIND-CAND]
        {Find\\Functional\\Blocks};

        \node[main node, tall] (BuildDGraph) [right=of FindFuncBlocks] {Build\\Delta\\Graph};

        \node[main node, tall] (MinSubGraph) [right= of BuildDGraph] { Minimum\\Induced\\Subgraphs};

        \node[main node, tall] (Simulate) [right= of MinSubGraph] {Simulation};

        \node[main node, tall] (Finalize) [right= of Simulate] {Output};

        \node[draw=none] (memvals1) [right=15pt of Finalize] {\texttt{(addr, value)}};
        \node[draw=none] (memvals2) [above=10pt of memvals1] {\texttt{(addr, value)}};
        \node[draw=none] (memvals4) [above=-2pt of memvals1] {\texttt{(addr, value)}};

        \node[draw=none] (memdots)  [below=2pt of memvals1] {$\dots$};
        \node[draw=none] (memvals3) [below=10pt of memvals1] {\texttt{(addr, value)}};

        \draw [decorate, decoration={brace, amplitude=3pt, mirror}, transform canvas={xshift=-30pt}]
        (memvals2.north) -- (memvals3.south);

        \draw[myarrow] (BIN-E)          -- (BIN-FE);
        \draw[myarrow] (SPL-E)          -- (SPL-FE);
        \draw[myarrow] (BIN-FE) -- ([yshift=26pt] FIND-CAND.west);
        \draw[myarrow] (SPL-FE) -- ([yshift=-26pt] FIND-CAND.west);
        \draw[myarrow] (FIND-CAND) -- (FindFuncBlocks);
        \draw[myarrow, transform canvas={yshift=20pt}] (FIND-CAND) -- (FindFuncBlocks);
        \draw[myarrow, transform canvas={yshift=-20pt}] (FIND-CAND) -- (FindFuncBlocks);
        \draw[myarrow, transform canvas={yshift=20pt}] (FindFuncBlocks) -- (BuildDGraph);
        \draw[myarrow, transform canvas={yshift=-20pt}] (FindFuncBlocks) -- (BuildDGraph);
        \draw[myarrow] (BuildDGraph) -- (MinSubGraph);
        \draw[myarrow] (Finalize) -- (memvals1);
        \draw[myarrow] (MinSubGraph) -- (Simulate.west);
        \draw[myarrow] (Simulate) --  (Finalize);

        { %
            \coordinate(pt1) at ($(Simulate) + (1.05cm, -1.75cm)$);
            \coordinate(pt2) at ($(Simulate) + (1.05cm, 1.75cm)$);
            \draw[myarrow,red,draw=red] ([xshift=10pt]Simulate.south) |- (pt1) |- (pt2) -| ([xshift=10pt]Simulate.north);
        }

        { %
            \coordinate(pt1) at ($(BuildDGraph) + (0.75cm, -1.75cm)$);
            \coordinate(pt2) at ($(BuildDGraph) + (0.75cm, 1.75cm)$);
            \draw[myarrow,red,draw=red] ([xshift=8pt]BuildDGraph.south) |- (pt1) |- (pt2) -| ([xshift=10pt]BuildDGraph.north);
        }

        {
            \coordinate(pt1) at ($(Simulate)!0.5!(MinSubGraph) + (0,-1.75cm)$);
            \draw[myarrow,red,draw=red] ([xshift=-10pt]Simulate.south) |- (pt1) -| ([xshift=10pt]MinSubGraph.south);
        }

        {
            \coordinate(pt1) at ($(BuildDGraph)!0.5!(MinSubGraph) + (0,-2cm)$);
            \draw[myarrow,red,draw=red] ([xshift=-10pt]MinSubGraph.south) |- (pt1) -| (BuildDGraph.south);
        }

        {
            \coordinate(pt1) at ($(FindFuncBlocks)!0.5!(BuildDGraph) + (0,-2cm)$);
            \draw[myarrow,red,draw=red] ([xshift=-8pt]BuildDGraph.south) |- (pt1) -| (FindFuncBlocks.south);
        }

        \node[draw=none] (labelN) [above=10pt of MinSubGraph] {$N$};
        \draw[myarrow] (labelN) to (MinSubGraph);

        \node[draw=none] (labelK) [above=10pt of Simulate] {$K$};
        \draw[myarrow] (labelK) to (Simulate);

        \node[draw=none] (labelP) [above=10pt of BuildDGraph] {$P$};
        \draw[myarrow] (labelP) to (BuildDGraph);

        \node[draw=none] (labelL) [above=10pt of BuildDGraph, transform canvas={xshift=-10pt}] {$L$};
        \draw[myarrow, transform canvas={xshift=-10pt}] (labelL) to (BuildDGraph);

        \node[draw=none] [above right=-10pt and 0pt  of BIN-FE] {$CFG_A$};
        \node[draw=none] [above right=-10pt and 5pt  of SPL-FE] {$IR$};
        \node[draw=none] [above=2pt of $(FIND-CAND)!0.5!(FindFuncBlocks)$, transform canvas={yshift=20pt}] {$R_G$};
        \node[draw=none] [above=2pt of $(FIND-CAND)!0.5!(FindFuncBlocks)$] {$V_G$};
        \node[draw=none] [above=2pt of $(FIND-CAND)!0.5!(FindFuncBlocks)$, transform canvas={yshift=-20pt}] {$C_B$};
        \node[draw=none] [above=20pt of $(FindFuncBlocks.east)!0.5!(BuildDGraph.west)$] {$F_B$};
        \node[draw=none] [below=2pt of $(FindFuncBlocks.east)!0.5!(BuildDGraph.west)$] {$M_{Adj}$};

        \node[draw=none] [above=2pt of $(BuildDGraph.east)!0.5!(MinSubGraph.west)$] {$\delta G$};
        \node[draw=none] [above=2pt of $(MinSubGraph.east)!0.5!(Simulate.west)$] {$H_k$};
        \node[draw=none] [above=2pt of $(Simulate.east)!0.5!(Finalize.west)$] {$C_w$};

    \end{tikzpicture}
}
  \caption{High level overview of the \sysname implementation. The red arrows indicate the
  iterative process upon failure. 
$CFG_A$: CFG with basic block abstractions added,
$IR$: Compiled SPL payload
$R_G$: Register mapping graph,
$V_G$: All variable mapping graphs,
$C_B$: Set of candidate blocks,
$F_B$: Set of functional blocks,
$M_{Adj}$: Adjacency matrix of SPL payload,
$\delta G$: Delta graph,
$H_k$: Induced subgraph,
$C_w$: Constraint set.
$L$: Maximum length of continuous dispatcher blocks,
$P$: Upper bound on payload ``shuffles'',
$N$: Upper bound on minimum induced subgraphs,
$K$: Upper bound on shortest paths for dispathers.
  }
  \label{fig:overview}
\end{figure*}

Our open source prototype, \sysname, is implemented in Python and consists of 
approximately $14,\!000$ lines of code. The current prototype focuses on x64
binaries, we leave the (straightforward) extension to other architectures such
as x86 or ARM as future work. 
\sysname requires three distinct inputs:
\begin{itemize}
  \item The exploit payload expressed in SPL,
  \item The vulnerable application on top of which the payload runs,
  \item The entry point in the vulnerable application, which is a location 
  that the program reaches naturally and occurs after all AWPs
  have been completed.
\end{itemize}

The output of \sysname is a sequence of \((address, value, size)\) tuples 
that describe how the memory should be modified during the state modification phase 
(\autoref{sec:threat}) to execute the payload.
Optionally, it may also generate some additional \((stream, value, size)\) 
tuples that describe what additional input should be given on any potentially 
open ``streams'' (file descriptors, sockets, stdin) that the attacker controls 
during the execution of the payload.

A high level overview of \sysname is shown in \autoref{fig:overview}.
Our algorithm is {\it iterative}; that is, in case of a failure, the red 
arrows, indicate which module is executed next.

\subsection{Binary Frontend} \label{sec:binfe}

The Binary Frontend uses angr~\cite{angr} to lift the target binary into the
VEX intermediate representation to expose the application's CFG.
Operating directly on basic blocks is cumbersome and heavily dependent on 
the Application Binary Interface (ABI). Instead, we translate each
basic block into a \emph{block constraint summary}.
Abstraction leverages symbolic execution~\cite{sym}
to ``summarize'' the basic block into a set of constraints encoding changes 
in registers and memory, and any potential system, library call, or conditional 
jump at the end of the block -- generally any \emph{effect} that this block 
has on the program's state.
\sysname  executes each basic block in an isolated environment,
	where every action (such as accesses to registers or memory) is monitored.
Therefore, instead of working with the instructions of each basic block, \sysname
utilizes its abstraction for all operations.
The abstraction information for every basic block is added to the CFG, resulting 
in $CFG_A$.

\subsection{SPL Frontend} \label{sec:splfe}

The SPL Front end translates the exploit payload into a graph-based Intermediate
Representation (IR) for further processing. %
To increase the flexibility of the mapping process, statements in a sequence
can be executed out-of-order. For each statement sequence we build a
\emph{dependence graph} based on a customized version of Kahn's
topological sorting algorithm~\cite{kahn},
to infer all groups of independent statements.
Independent statements in a subsequence are then turned into a set of statements
which can be executed out-of-order.
This results in a set of equivalent payloads that are permutations
of the original. Our goal is to find a solution for \emph{any} of them.

\subsection{Locating candidate block sets} \label{sec:mark}

\setcellgapes{1pt} %
\begin{table*}[ht]
\makegapedcells
\resizebox{\textwidth}{!}{
\centering
\begin{tabular}{|l|l|l|l|l|l|l|}
\hline
\multicolumn{1}{|c|}{\textbf{Statement}}   &
\multicolumn{1}{c|}{\textbf{Form}}        &
\multicolumn{1}{c|}{\textbf{Abstraction}} &
\multicolumn{3}{ c|}{\textbf{Actions}}     &
\multicolumn{1}{c|}{\textbf{Example}} \\ \hline

\multirow{4}{*}{\textit{\begin{tabular}[c]{@{}l@{}}Register Assignment\end{tabular}}} 
& \multirow{2}{*}{$\underline{\hspace{.3cm}}r_{\alpha} = C$} 
& $reg_{\gamma} \leftarrow C$                    
& \multirow{5}{*}{$R_G\cup \big\{(\underline{\hspace{.3cm}}r_{\alpha},reg_{\gamma})\big\}$} 
& \multicolumn{2}{c|}{--}
& \texttt{movzx rax, 7h} \\ \cline{3-3} \cline{5-7} 
                                                                                        &
& $reg_{\gamma} \leftarrow *A$                    
&
& \multicolumn{2}{c|}{$D_M \cup \{A\}$}                         
& \texttt{mov rax, ds:fd} \\ \cline{2-3} \cline{5-7} 
                                                                                       
& \multirow{2}{*}{$\underline{\hspace{.3cm}}r_{\alpha} = \&V$} 
& $reg_{\gamma} \leftarrow C, \; C\!\in\! R\!\land\!W$
& 
& \multirow{2}{*}{$V^{\alpha\gamma}_G \cup \big\{(V,A)\big\}$} 
& \multicolumn{1}{c|}{--}
& \texttt{lea rcx, [rsp+20h]} \\ \cline{3-3} \cline{6-7} 

&
& $reg_{\gamma} \leftarrow *A$
&
&
& $D_M \cup \{A\}$ 
& \texttt{mov rdx, [rsi+18h]} \\ \hline

\textit{Register Modification}                                                          & $\underline{\hspace{.3cm}}r_{\alpha}\; \odot\!= C$
& $reg_{\gamma} \leftarrow reg_{\gamma} \odot C$
& \multicolumn{3}{l|}{$R_G\cup\big\{(\underline{\hspace{.3cm}}r_{\alpha},reg_{\gamma})\big\}$} 
& \texttt{dec rsi} \\ \hline

\textit{Memory Read}                                                                    
& $\underline{\hspace{.3cm}}r_{\alpha} = * \underline{\hspace{.3cm}}r_{\beta}$
& $reg_{\gamma} \leftarrow *reg_{\delta}$
& \multicolumn{3}{l|}{\multirow{2}{*}
{$R_G\cup\big\{(\underline{\hspace{.3cm}}r_{\alpha},reg_{\gamma}),
\;(\underline{\hspace{.3cm}}r_{\beta},reg_{\delta})\big\}$}}
& \texttt{mov rax, [rbx]} \\ \cline{1-3} \cline{7-7}

\textit{Memory Write}
& $*\underline{\hspace{.3cm}}r_{\alpha} = \underline{\hspace{.3cm}}r_{\beta}$
& $*reg_{\gamma} \leftarrow reg_{\delta}$
& \multicolumn{3}{l|}{}
& \texttt{mov [rax], [rbx]} \\ \hline

\textit{Call}
& $call(\underline{\hspace{.3cm}}r_{\alpha},\;\underline{\hspace{.3cm}}r_{\beta},\;...)$
& \texttt{Ijk_Call} to $call$
& \multicolumn{3}{l|}{$R_G \cap\big\{(\underline{\hspace{.3cm}}r_{\alpha}, \%rdi), 
\;(\underline{\hspace{.3cm}}r_{\beta}, \%rsi),\;...\big\}$}
& \texttt{call execve} \\[1ex] \hline

\textit{Conditional Jump}
& \begin{tabular}[l]{@{}l@{}}
$if\;(\underline{\hspace{.3cm}}r_{\alpha}\;\odot\!=\;C)$ \\ 
$\;\;\;\;goto \;LOC$
\end{tabular}

& \begin{tabular}[l]{@{}l@{}}
\texttt{Ijk_Boring} $\land$ \\
$condition = reg_{\gamma} \odot C$
\end{tabular}
& \multicolumn{3}{l|}{$R_G\cup\big\{(\underline{\hspace{.3cm}}r_{\alpha},reg_{\gamma})\big\}$} 
& \begin{tabular}[l]{@{}l@{}}
\texttt{test rax, rax} \\
\texttt{jnz LOOP} 
\end{tabular} \\ \hline
\end{tabular}
}
\caption{Semantic matching of SPL statements to basic blocks.
\textbf{Abstraction} indicates the requirements that the basic block 
abstraction needs to have to match the SPL statement in the \textbf{Form}. 
Upon a match, the appropriate \textbf{Actions} are taken.
$\underline{\hspace{.3cm}}r_{\alpha}$, $\underline{\hspace{.3cm}}r_{\beta}$:
Virtual registers,
$reg_{\gamma}$, $reg_{\delta}$: Hardware registers,
$C$: Constant value,
$V$: SPL variable,
$A$: Memory address,
$R_G$: Register mapping graph, 
$V_G$: Variable mapping graph, 
$D_M$: Dereferenced Addresses Set,
{\tt Ijk_Call}: A call to an address,
{\tt Ijk_Boring}: A normal jump to an address.
}
\label{tab:match}
\vspace{-2em}
\end{table*}

SPL is a high level language that hides the underlying ABI. Therefore,
\sysname looks for potential ways to ``map'' the SPL environment to the 
underlying ABI. The key insight in this step is to find all possible ways 
to map the individual elements from the SPL environment to the ABI (though
{\it candidate blocks}) and then iteratively selecting valid subsets from the 
ABI to ``simulate'' the environment of the SPL payload.
 
Once the $CFG_A$ and the IR are generated, \sysname searches for and marks {\it
candidate} basic blocks, as described in \autoref{sec:functional}.  For a block 
to be a candidate, it must ``semantically match'' with one (or more) payload 
statements. \autoref{tab:match} shows the matching rules.  Note that variable 
assignments, unconditional jumps, and returns do not require a basic block and 
therefore are excluded from the search.

All statements that assign or modify registers require the basic block to
apply the same operation on the same, as yet undetermined, hardware registers.
For function calls, the requirement for the basic block is to invoke the 
same call, either as a system call or as a library call
(if the arguments are different, the block is clobbering).
Note that the calling convention exposes the register mapping.

Upon a successful matching, \sysname builds the following data structures:

\begin{itemize}
\item $R_G$, the {\it Register Mapping Graph} which is a bipartite undirected 
	  graph. 
	  The nodes in the two sets represent the virtual and hardware 
	  registers respectively. The edges represent potential associations 
	  between virtual and hardware registers.

\item $V_G$, the {\it Variable Mapping Graph}, which is very similar to $R_G$,
	  but instead associates payload variables to underlying memory
	  addresses. $V_G$ is unique for every edge in $R_G$ i.e.:
		\begin{align*}
			\forall (\underline{\hspace{.3cm}}r_{\alpha}, reg_{\gamma}) \in R_G 
			\; \exists ! \, V_G^{\alpha\gamma}
		\end{align*}

\item $D_M$, the {\it Memory Dereference Set}, which has all memory addresses
	  that are dereferenced and their values are loaded into registers.
	  Those addresses can be symbolic expressions (e.g., \texttt{[rbx + 
	  rdx*8]}), and therefore we do not know the concrete address
	  they point to until execution reaches them (see \autoref{sec:simu}).

\end{itemize}

After this step, each SPL statement has a {\it set} of candidate blocks. 
Note that a basic block can be candidate for multiple statements. If 
for some statement there are no candidate blocks, the algorithm halts and 
reports that the program cannot be synthesized.

\subsection{Identifying functional block sets} \label{sec:remark}

After determining the set of candidate blocks, $C_B$, \sysname
\emph{iteratively} identifies, for each SPL statement, which 
candidate blocks can serve as functional blocks, i.e., the blocks that perform 
the operations.
This step determines for each candidate block if there is a resource
mapping that satisfies the block's constraints.

\sysname identifies the {\it concrete} set of hardware registers and memory
addresses that execute the desired statement.  A successful mapping identifies
candidate blocks that can serve as functional blocks.

To find the hardware-to-virtual register association, \sysname searches for a
\emph{maximum bipartite matching}~\cite{algo} in $R_G$. If such a mapping does
not exist, the algorithm halts.
The selected edges indicate the set of $V_G$ graphs that are used to find the
memory mapping, i.e., the variable-to-address association (see 
\autoref{sec:mark}, there can be a $V_G$
for every edge in $R_G$). Then for every $V_G$ the algorithm repeats the same
process to find another maximum bipartite matching.

This step determines, for each statement, which concrete registers and
memory addresses are reserved. Merging this information with the set of candidate blocks
constructs each block's SPL state, enabling the removal of candidate blocks that are
unsatisfiable.

However, there may be multiple candidate blocks for each SPL statement, and
thus the maximum bipartite match may not be unique. The algorithm
enumerates \emph{all} maximum bipartite matches~\cite{allmatch},
trying them one by one. If no match leads to a solution, the
algorithm halts.

\subsection{Selecting functional blocks} \label{sec:selfunc}

Given the functional block set $F_B$, this step searches for a subset
that executes all payload statements.
The goal is to select \emph{exactly} one functional block for every IR
statement and find dispatcher blocks to chain them together. \sysname
builds the {\it delta graph} $\delta G$, described in
\autoref{sec:delta}.

Once the delta graph is generated, this step locates the {\it 
minimum} (in terms of total edge weight) {\it induced subgraph}, $H_{k_0}$,
that contains the {\em complete} set of functional blocks to 
execute the SPL payload. If $H_{k_0}$, does not result
in a solution, the algorithm tries the next minimum induced subgraph,
$H_{k_1}$, until a solution is found or a limit is reached. 

If the resulting delta graph does not lead to a solution, this step
``shuffles'' out-of-order payload statements, see
\autoref{sec:splfe}, and builds a new delta graph.
Note that the number of different permutations may be exponential.
Therefore, our algorithm sets an {\it upper bound} $P$ on the number of tried
permutations.

Each permutation results in a different yet semantically equivalent SPL payload,
so the CFG of the payload (called {\it Adjacency Matrix}, $M_{Adj}$)
needs to be recalculated.

\subsection{Discovering dispatcher blocks} \label{sec:simu}

The simulation phase takes the individual functional blocks (contained in
the minimum induced subgraph $H_{k_i}$) and tries to find the appropriate 
dispatcher blocks to compose the BOP gadgets.
It returns a set of memory assignments for the corresponding dispatcher blocks,
or an error indicating un-satisfiable constraints for the dispatchers.

\sysname is called to find a dispatcher path for {\it every} edge in the 
minimum induced subgraph. That is, we need to simulate every control flow
transfer in the adjacency matrix, $M_{Adj}$ of the SPL payload.
However, dispatchers are {\it built} on the prior set of BOP gadgets and their
impact on the binary's execution state so far, so BOP 
gadgets must be stitched with the respect to the program's current 
flow originating from the entry point.

Finding dispatcher blocks relies on concolic execution. Our algorithm
utilizes functional block proximity as a metric for dispatcher path
quality.  However, it cannot predict which constraints will take
exponential time to solve (in practice we set a timeout). Therefore
concolic execution selects the $K$ shortest dispatcher paths relative
to the current BOP chain, and tries them in order until one produces a
set of satisfiable constraints.  It turns that this metric works well
in practice even for small values of $K$ (e.g., $8$).
This is similar to the {\it k-shortest path}~\cite{KSP} algorithm used
for the delta graph.

When simulation starts it also initializes any SPL variables at the locations
that are reserved during the variable mapping (\autoref{sec:remark}). These
addresses are marked as \emph{immutable}, so any unintended modification raises
an exception which stops this iteration. %

In \autoref{tab:match}, we introduce the set of {\it Dereferenced Addresses},
$D_M$, which is the set of memory addresses whose contents are loaded into
registers. Simulation cannot obtain the exact location of a symbolic address
(e.g., \texttt{[rax + 4]}) until the block is executed and the register has a
concrete value.  Before simulation reaches a functional block, it concretizes
any symbolic addresses from $D_M$ and initializes the memory cell accordingly.
If that memory cell has already been set, any initialization \emph{prior} to the
entry point cannot persist.  That is, \sysname cannot leverage an AWP to
initialize this memory cell and the iteration fails.
If a memory cell has been used in the constraints, its concretization can make
constraints unsatisfiable and the iteration may fail.

Simulation traverses the minimum induced subgraph, and {\it incrementally} 
extends the SPL state from one BOP gadget to the next, ensuring that
newly added constraints remain satisfiable.
When encountering a conditional statement (i.e., a functional block has
two outgoing edges), \sysname \textit{clones} the current state and continues 
building the trace for both paths independently, in the same way that a 
symbolic execution engine handles conditional statements. 
When a path reaches a functional block that was already visited, it gracefully 
terminates.
At the end, we collect all those states and check whether the constraints of 
all these paths are satisfied or not. If so, we have a 
solution.

\subsection{Synthesizing exploits} \label{sec:extr}

\newcolumntype{C}[1]{>{\centering\let\newline\\\arraybackslash\hspace{0pt}}m{#1}}
\newcolumntype{M}[1]{>{\centering\arraybackslash}m{#1}}

\begin{table*}[ht]
\centering
\begin{adjustbox}{width=1\textwidth}
\begin{tabular}{|l|l|l|r|r|r|r|r|r|r|r|r|r|}
\hline
\multicolumn{3}{|c|}{\textbf{Vulnerable Application}} & 
\multicolumn{2}{c|}{\textbf{CFG}} & 
\multicolumn{1}{c|}{\multirow{2}{*}{
   \textbf{\begin{tabular}[c]{@{}l@{}}Time\\(m:s)\end{tabular}}}} & 
\multicolumn{7}{c|}{\textbf{Total number of functional blocks}} \\ 
\cline{1-5} \cline{7-13} 

\multicolumn{1}{|c|}{\textbf{Program}} & 
\multicolumn{1}{c|}{\textbf{Vulnerability}} &
\multicolumn{1}{c|}{\textbf{Prim.}} &
\multicolumn{1}{c|}{\textbf{Nodes}}& \multicolumn{1}{l|}{\textbf{Edges} }& 
\multicolumn{1}{c|}{}  &
\multicolumn{1}{C{1.2cm}|}{\textbf{RegSet}} &
\multicolumn{1}{C{1.2cm}|}{\textbf{RegMod}} &
\multicolumn{1}{C{1.2cm}|}{\textbf{MemRd}}  &
\multicolumn{1}{C{1.2cm}|}{\textbf{MemWr}}  &
\multicolumn{1}{C{1.2cm}|}{\textbf{Call}}   &
\multicolumn{1}{C{1.2cm}|}{\textbf{Cond}}   &
\multicolumn{1}{C{1.2cm}|}{\textbf{Total}} \\ \hline

ProFTPd & CVE-2006-5815~\cite{proFTPdCVE} & AW & 27,087 & 49,862 & 10:08
& 40,143 & 387 & 1,592 & 199 & 77 & 3,029 & 45,427 \\ \hline
                                 
nginx   & CVE-2013-2028~\cite{nginxCVE}   & AW & 24,169 & 44,645 & 12:36
& 31,497 & 1,168 & 1,522 & 279 & 35 & 3375 & 37,876 \\ \hline

sudo    & CVE-2012-0809~\cite{sudoCVE}    & FMS & 3,399 & 6,267 & 01:14
& 5,162 & 26 & 157 & 18 & 45 & 307 & 5715 \\ \hline

orzhttpd & BugtraqID 41956~\cite{orzhttpdCVE} & FMS & 1,354 & 2,163 & 00:27
& 2,317 & 9 & 39 & 8 & 11 & 89 & 2473 \\ \hline

wuftdp  & CVE-2000-0573~\cite{wuftpdCVE} & FMS & 8,899 & 17,092 & 03:22
& 14,101 & 62 & 274 & 11 & 94 & 921 & 15,463 \\ \hline

nullhttpd & CVE-2002-1496~\cite{nullhttpdCVE} & AW & 1,488 & 2,701 & 00:27
& 2,327 & 77 & 54 & 7 & 19 & 125 & 2,609 \\ \hline

opensshd & CVE-2001-0144~\cite{opensshCVE} & AW & 6,688 & 12,487 & 01:53
& 8,800 & 98 & 214 & 19 & 63 & 558 & 9,752 \\ \hline

wireshark & CVE-2014-2299~\cite{wiresharkCVE} & AW & 74,186 & 162,111 & 29:41
& 12,4053 & 639 & 1,736 & 193 & 100 & 4555 & 131276 \\ \hline

apache & CVE-2006-3747~\cite{apacheCVE} & AW & 18,790 & 34,205 & 10:22
& 33,615 & 212 & 490 & 66 & 127 & 1,768 & 36,278 \\ \hline

smbclient & CVE-2009-1886~\cite{smbclientCVE} & FMS & 166,081 & 351,309 & 82:25
& 265,980 & 1,481 & 6,791 & 951 & 119 & 28,705 & 304,027 \\ \hline
\end{tabular}
\end{adjustbox}
\caption{\textbf{Vulnerable applications}. The {\it Prim.} column indicates 
the primitive type ({\tt AW} = Arbitrary Write, {\tt FMS} = ForMat String). {\it Time} is 
the amount of time needed to generate the abstractions
for every basic block. {\it Functional blocks} show the total number for each 
of the statements ({\it RegSet} = Register Assignments, {\it RegMod} = Register
Modifications, {\it MemRd} = Memory Load, {\it MemWr} = Memory Store, 
{\it Call} = system/library calls, {\it Cond} = Conditional Jumps). 
Note that the number of call statements is small
because we are targeting a predefined set of calls. Also note that MemRd 
statements are a subset of RegSet statements.}
\label{tab:applications}
\vspace{-1.5em}
\end{table*}

\begin{table}[]
\centering
\begin{adjustbox}{width=1\linewidth}
\begin{tabular}{l|l|r|c}

\multicolumn{1}{c}{\textbf{Payload}} & \multicolumn{1}{c}{\textbf{Description}} &
\multicolumn{1}{c}{\textbf{$|S|$}} & \textbf{flat?} \\ \hline

\textit{regset4} & Initialize 4 registers with arbitrary values
	& 4 	& \cmark  \\ 
\textit{regref4} & Initialize 4 registers with pointers to arbitrary memory
	& 8 & \cmark  \\
\textit{regset5} & Initialize 5 registers with arbitrary values
    & 5 & \cmark  \\ 
\textit{regref5} & Initialize 5 registers with pointers to arbitrary memory
    & 10 & \cmark  \\ 
\textit{regmod}  & Initialize a register with an arbitrary value and modify it
	& 3 & \cmark  \\ 
\textit{memrd}   & Read from arbitrary memory
	& 4 & \cmark  \\ 
\textit{memwr}   & Write to arbitrary memory
    & 5 & \cmark  \\ 
\textit{print}   & Display a message to stdout using \texttt{write}
	& 6 & \cmark  \\ 	
\textit{execve}  & Spawn a shell through \texttt{execve}
	& 6 & \cmark  \\ 
\textit{abloop}  & Perform an arbitrarily long bounded loop utilizing regmod
    & 2 & \xmark  \\
\textit{infloop} &Perform an infinite loop that sets a register in its body
    & 2 & \xmark  \\
\textit{ifelse}  & An if-else condition based on a register comparison
    & 7 & \xmark  \\
\textit{loop}    & Conditional loop with register modification
    & 4 & \xmark  \\
\end{tabular}
\end{adjustbox}

\caption{SPL payloads. Each payload consists of $|S|$ statements. Payloads
that produce {\it flat} delta graphs (i.e., have no jump statements), are 
marked with \cmark. \textit{memwr} payload modifies program 
memory on the fly, thus preserving the Turing completeness of SPL 
(recall from \autoref{sec:threat} that AWP/ARP-based state modification is no
longer allowed).}
\label{tab:payloads}
\vspace{-3em}
\end{table}

If the simulation module returns a solution, the final step is to encode the
execution trace as a set of memory writes in the target binary.
The constraint set $C_w$ collected during simulation reveals a memory layout 
that leads to a flow across functional blocks according to the 
minimum induced subgraph.
Concretizing the constraints for all participating conditional 
variables at the \emph{end} of the simulation can result in incorrect
solutions. Consider the following case:%

\begin{minipage}[t]{\textwidth}
\centering
\begin{minted}[fontsize=\small, mathescape=true, framesep=1mm, baselinestretch=1.0,]{c}
	a = input();
	if (a > 10 && a < 20) {
		a = 0;
		/* target block */
	}
\end{minted}
\end{minipage}

The symbolic execution engine concretizes the symbolic variable assigned to 
\texttt{a} upon assignment. When execution reaches ``target block'',
\texttt{a} is $0$, which is contradicts the precondition to reach the
target block. Hence, \sysname needs to resolve the constraints \emph{during}
(i.e., {\it on the fly}), rather than at the end of the simulation.
Therefore, constraints are solved inline in the simulation. \sysname carefully 
monitors all variables and concretizes them at the ``right'' moment, 
just before they get overwritten.
More specifically, memory locations that are accessed for first time, are 
assigned a symbolic variable. Whenever a memory write occurs, \sysname checks 
whether the initial symbolic variable still exists %
in the new symbolic expression. If not, \sysname concretizes it, adding the 
concretized value to the set of memory writes.

There are also some symbolic variables that do {\it not} participate in the
constraints, but are used as pointers. These variables are
concretized to point to a writable location to avoid 
segmentation faults outside of the simulation environment.

Finally, it is possible for registers or external symbolic variables (e.g., 
data from stdin, sockets or file descriptors) to be part of the constraints.
\sysname executes a similar translation for the registers and any
external input, as these are inputs to the program that are usually also 
controlled by the attacker.

\section{Evaluation} \label{sec:eval}

\newcommand{\s} {\scriptstyle\;}
\begin{table*}[ht]
\centering
\begin{adjustbox}{width=1\textwidth}
\begin{tabular}{|l|M{1cm}|M{1cm}|M{1cm}|M{1cm}|M{1cm}|M{1cm}|M{1cm}
                  |M{1cm}|M{1cm}|M{1cm}|M{1cm}|M{1cm}|M{1cm}|}
\hline
\multirow{2}{*}{\textbf{Program}} & \multicolumn{13}{c|}{\textbf{SPL payload}}
\\ \cline{2-14} &
\textit{regset4} & \textit{regref4} & \textit{regset5} & \textit{regref5} &
\textit{regmod}  & \textit{memrd}   & \textit{memwr}   & \textit{print}   &
\textit{execve}  & \textit{abloop}  & \textit{infloop}& \textit{ifelse}  &
\textit{loop} \\ \hline
ProFTPd   &\ok &\ok &\ok &\ok &\ok &\ok &\ok &\ok$\s 32$ &\xA
		  &\ok$\s 128+$ &\ok$\s \;\;\;\;\infty$ &\ok &\ok$\s \;\;\;3$ \\ \hline
		  
nginx     &\ok &\ok &\ok &\ok &\ok &\ok &\ok &\xD &\ok
		  &\ok $\s 128+$ &\ok $\s \;\;\;\;\infty$ &\ok &\ok $\s 128$ \\ \hline
		  
sudo      &\ok &\ok &\ok &\ok &\ok &\ok &\ok &\ok &\ok
		  &\xD &\ok $\s 128+$ &\xD &\xD \\ \hline

orzhttpd  &\ok &\ok &\ok &\ok &\ok &\ok &\ok &\xD &\xA 
		  &\xD &\ok $\s 128+$ &\xD &\xC \\ \hline

wuftdp    &\ok &\ok &\ok &\ok &\ok &\ok &\ok &\ok &\xA 
		  &\ok $\s 128+$ &\ok $\s 128+$ &\xD &\xC \\ \hline

nullhttpd &\ok &\ok &\ok &\ok &\ok &\ok &\xC &\xC &\ok 
		  &\ok $\s \;\;\;30$ &\ok $\s \;\;\;\;\infty$ &\xD &\xC \\ \hline

opensshd  &\ok &\ok &\ok &\ok &\ok &\ok &\xD &\xD &\xD 
		  &\ok $\s \;\;512$ &\ok $\s 128+$ &\ok &\ok $\s \;\;99$ \\ \hline

wireshark &\ok &\ok &\ok &\ok &\ok &\ok &\ok &\ok $\s 4$ &\xA 
		  &\ok $\s 128+$ &\ok $\s \;\;\;\;7$ &\ok &\ok $\s \;\;\;8$ \\ \hline

apache    &\ok &\ok &\ok &\ok &\ok &\ok &\ok &\xD &\xD 
		  &\ok $\s \;\;\;\;\infty$ &\ok $\s 128+$ &\ok &\xD \\ \hline

smbclient &\ok &\ok &\ok &\ok &\ok &\ok &\ok &\ok $\s 1$ &\xA 
		  &\ok $\s 1057$ &\ok $\s 128+$ &\ok &\ok $\s 256$\\ \hline

\end{tabular}
\end{adjustbox}

\caption{Feasibility of executing various SPL payloads for each of the vulnerable
applications. An \ok means that the SPL payload was successfully executed on the
target binary while a \xmark indicates a failure, with the subscript denoting
the type of failure (\xA = Not enough candidate blocks, \xB = No valid
register/variable mappings, \xC = No valid paths between functional blocks and
\xD = Un-satisfiable constraints or solver timeout). Note that in the first two
cases (\xA and \xB), we know that there is \emph{no} solution while, in the last
two (\xC and \xD), a solution might exists, but \sysname cannot find it, either
due to over-approximation or timeouts. The numbers next to the \ok in {\it
abloop}, {\it infloop}, and {\it loop} columns indicate the maximum number of
iterations. The number next to the {\it print} column indicates the number of
character successfully printed to the stdout.  \vspace{-15pt} }
\label{tab:results} \end{table*}

To evaluate \sysname, we leverage a set of $10$ applications with known 
memory corruption CVEs, listed in \autoref{tab:applications}. These CVEs
correspond to arbitrary memory writes~\cite{dop, dop-gadget, cfb}, fulfilling
our AWP primitive requirement.
\autoref{tab:applications} contains the total number of all functional blocks 
for each application. Although there are many functional blocks,
the difficulty of finding stitchable dispatcher blocks
makes a significant fraction of them unusable.

Basic block abstraction is a time consuming process -- especially for 
applications with large CFGs -- but these results may be reused 
across iterations.
Thus, as a performance optimization, \sysname caches the resulting
abstractions of the Binary Frontend (\autoref{fig:overview}) to a file and
loads them for each search, thus avoiding the startup overhead listed in 
\autoref{tab:applications}.

To demonstrate the effectiveness of our algorithm, we chose a set of 
$13$ representative SPL payloads~\footnote{Results depend on the SPL payloads
and the vulnerable applications. We chose the SPL payloads to showcase all SPL
features, other payloads or combination of payloads are possible. We encourage
the reader to play with the open-source prototype.}
shown in \autoref{tab:payloads}. Our goal 
is  to ``map and run'' each of these payloads on top each of the vulnerable 
applications.
\autoref{tab:results} shows the results of running each payload.
\sysname successfully finds a mapping of memory writes 
to encode an SPL payload as a set of side effects executed on top 
of the applications for $105$ out of $130$ cases, approximately $81\%$.
In each case, the memory writes are sufficient to reconstruct the payload
execution by strictly following the CFG without violating a strict CFI policy
or stack integrity.

\autoref{tab:results} shows that
applications with large CFGs result in higher success rates, as they encapsulate
a ``richer'' set of BOP gadgets.
Achieving truly infinite loops is hard in practice, as most of the loops 
in our experiments involve some loop counter that is modified
in each iteration. This iterator
serves as an index to dereference an array. By falsifying the exit condition
through modifying loop variables (i.e., the loop becomes infinite), the program
eventually terminates with a segmentation fault,
as it tries to access memory outside of the current segment.
Therefore, even though the loop would run forever, an external
factor (segmentation fault) causes it to stop. \sysname aims to address this issue
by simulating the same loop multiple times. However, finding a truly infinite loop 
requires \sysname to simulate it an infinite number of times, which is infeasible. 
For some cases, we managed to verify that the accessed memory inside the loop is
bounded and therefore the solution truly is an infinite loop. Otherwise, the
loop is {\it arbitrarily bounded} with the upper bound set by an external factor.

For some payloads, \sysname was unable to find an exploit trace.
This is is either due to imprecision of our algorithm, or because no solution 
exists for the written SPL payload.
We can alleviate the first failure by increasing the upper bounds and the
timeouts in our configuration. Doing so, makes \sysname search more
exhaustively at the cost of search time.

The failure to find a solution exposes the \emph{limitations} of the vulnerable application.
This type of failure is due to the ``structure'' of the application's CFG, 
which prevents \sysname from finding a trace for an SPL payload. 
Hence, a solution may not exist due to one the following:

\begin{enumerate}
  \item There are not enough candidate blocks or functional blocks.
  \item There are no valid register / variable mappings.
  \item There are no valid paths 
  		between functional blocks.
  \item The constraints between blocks are unsatisfiable or symbolic execution
raised a timeout.
\end{enumerate}

For instance, if an application (e.g., ProFTPd) never invokes \texttt{execve}
then there are no candidate blocks for \texttt{execve} SPL satements.
Thus, we can infer from the {\it execve} column in \autoref{tab:results} that 
all applications with a \xA never invoke \texttt{execve}.

In \autoref{sec:threat} we mention that the determination of the entry point 
is part of the vulnerability discovery process. Therefore, \sysname assumes that
the entry point is given. Without having access to actual exploits (or crashes),
the locations of entry points are ambiguous.
Hence, we have selected arbitrary locations as the 
entry points. This allows \sysname to find payloads for the evaluation without
having access to \emph{concrete} exploits. In practice, \sysname would leverage
the given entry points as starting points.
We demonstrate several test cases where the entry points are precisely at 
the start of functions, deep in the Call Graph, to show the power of our approach. 
Orthogonally, we allow for vulnerabilities to exist in the middle
of a function.
In such situations, \sysname would set our entry point to the location \emph{after} 
the return of the function.

The lack of the exact entry point complicates the verification
of our solutions. We leverage a debugger to ``simulate'' the AWP and modify the
memory on the fly, as we reach the given entry point. 
We ensure as we step through our trace that we maintain the properties of
the SPL payload expressed. That is, blocks between the statements are
non-clobbering in terms of register allocation and memory assignment.

\section{Case Study: nginx}
We utilize a version of the nginx web server with a known memory corruption 
vulnerability~\cite{nginxCVE} that has been exploited in the wild  to further study \sysname.
When an HTTP header contains the ``Transfer-Encoding: chunked'' attribute, 
nginx fails to properly bounds check the received packet chunks, 
resulting in stack buffer overflow.
This buffer overflow~\cite{cfb} results in an 
arbitrary memory write, fulfilling the AWP requirement.
For our case study we select three of the most interesting payloads: spawning a
shell, an infinite loop, and a conditional branch.
\autoref{tab:performance} shows metrics collected during the \sysname execution for these cases.

\begin{table}[!h]
\centering
\begin{adjustbox}{width=.9\linewidth}
\begin{tabular}{|l|r|r|r|r|r|}
\hline
\multicolumn{1}{|c|}{\textbf{Payload}}       &
\multicolumn{1}{c|}{\textbf{Time}}     &
\multicolumn{1}{c|}{\textbf{$|C_B|$}}        &
\multicolumn{1}{c|}{\textbf{Mappings}} &
\multicolumn{1}{c|}{\textbf{$|\delta G|$}}  &
\multicolumn{1}{c|}{\textbf{$|H_k|$}} \\ \hline
execve   & 0m:55s & 10,407 & 142,355 & 1 & 1 \\ \hline
infloop  & 4m:45s & 9,909  & 14     & 1 & 1 \\ \hline
ifelse   & 1m:47s & 10,782 & 182    & 4 & 2 \\ \hline
\end{tabular}
\end{adjustbox}
\caption{Performance metrics (run on Ubuntu 64-bit with an
i7 processor) for \sysname on nginx.
	{\it Time} = time to synthesize exploit,
	$|C_B|$ = \# candidate blocks,
	{\it Mappings} = \# concrete register and variable mappings,
	{\it $|\delta G|$} = \# delta graphs created,
	{\it $|H_k|$} = \# of induced subgraphs tried.
    \vspace{-20pt}
}
\label{tab:performance}
\vspace{-1em}
\end{table}

\subsection{Spawning a shell}

Function \texttt{ngx_execute_proc} is invoked through a function pointer, with
the second argument (passed to \texttt{rsi}, according to x64 calling 
convention), being a \textbf{void} pointer that is interpreted as a 
\texttt{struct} to initialize all arguments of \texttt{execve}:

\begin{minipage}[t]{\textwidth}
\centering
\begin{minted}[fontsize=\small, mathescape=true, framesep=1mm, baselinestretch=1.0,]{text}
    mov   rbx, rsi
    mov   rdx, QWORD PTR [rsi+0x18]
    mov   rsi, QWORD PTR [rsi+0x10]
    mov   rdi, QWORD PTR [rbx]
    call  0x402500 <execve@plt>         
\end{minted}
\vspace{1em}
\end{minipage}

\sysname leverages this function to successfully synthesize the 
\texttt{execve} payload (shown on the right side of \autoref{tab:examples})
and generate a PoC exploit in less than a minute as shown in 
\autoref{tab:performance}.

Assuming that \texttt{rsi} points to some writable address $x$, \sysname
produces the following \((address, value, size)\) tuples:
$(\$y, \$x, 8)$, $(\$y+8h, 0, 8)$, $(\$x, /bin/sh, 8)$, $(\$x+10h, \$y, 8)$,
$(\$x+18h, 0, 8)$, were $\$y$ is a concrete writable addresses set by \sysname.

\subsection{Infinite loop}

\begin{figure}[t!]
\includegraphics[width=1\linewidth]{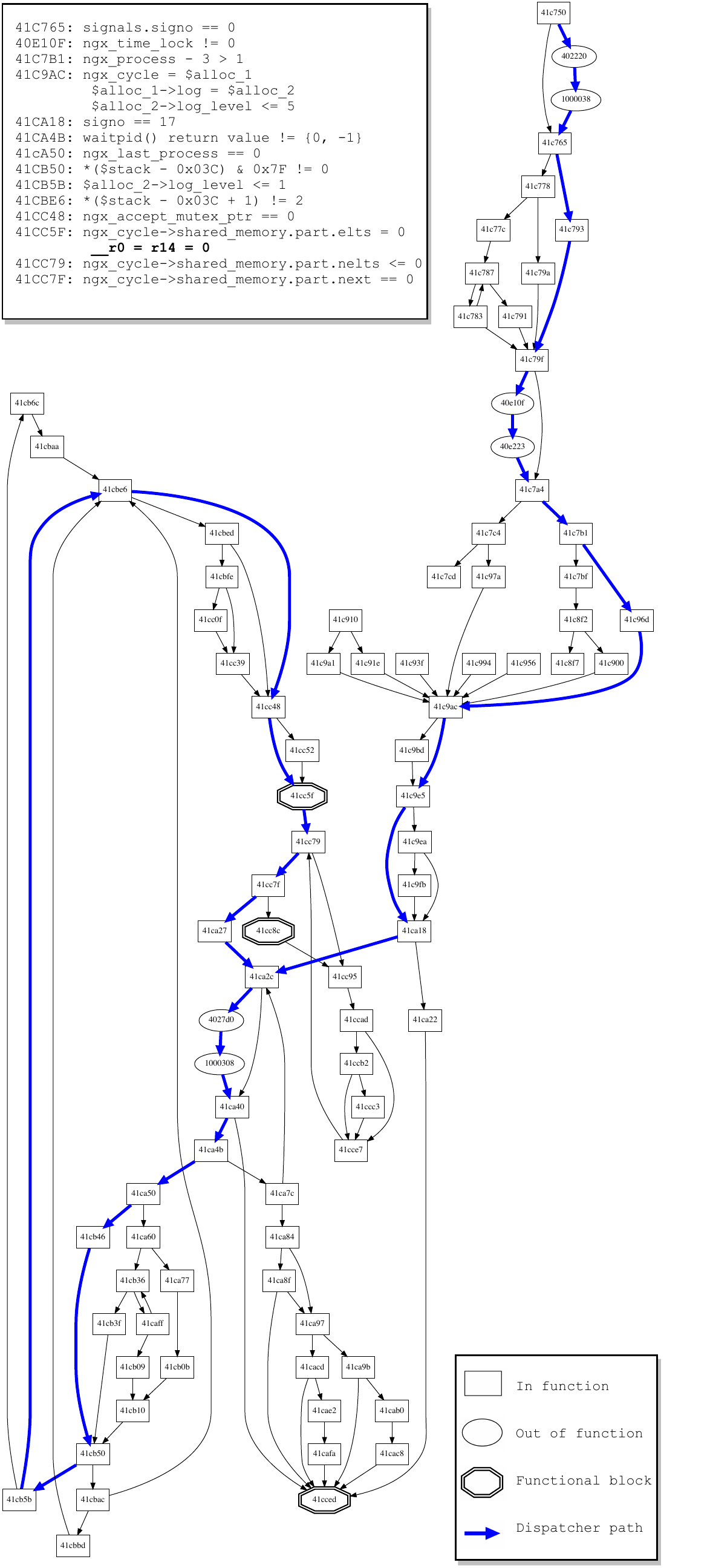}
\caption{CFG of nginx's \texttt{ngx_signal_handler} 
and payload for an infinite loop (blue arrow dispatcher blocks, octagons
functional blocks) with the entry point at the function start.  The top box
shows the memory layout initialization for this loop.  This graph was created 
by \sysname.
}\label{fig:infloop} \end{figure}

Here we present a payload that generates a trace that executes an infinite loop.
The {\it infloop} payload is a simple infinite loop that consists of only
two statements:

\begin{minipage}[t]{\textwidth}
\centering
\begin{minted}[fontsize=\small, mathescape=true, framesep=1mm, baselinestretch=1.0,]{c}
    void payload() {
      LOOP:
        __r1 = 0;
        goto LOOP;
    }
\end{minted}
\end{minipage}

We set the entry point at the beginning of \texttt{ngx_signal_handler}
function which is a signal handler that is invoked through a function
pointer. Hence, this point is reachable through control-flow
hijacking.  
The solution synthesized by \sysname is shown in
\autoref{fig:infloop}. The box on the top-left corner demonstrates
how the memory is initialized to satisfy the constraints. 

Virtual register \texttt{\_\_r0} was mapped to hardware register
\texttt{r14}, so \texttt{ngx_signal_handler} contains three candidate
blocks, marked as octagons. Exactly one of them is selected to be the
functional block while the others are avoided by the dispatcher
blocks. The dispatcher finds a path from the entry point to the first
functional block, and then finds a loop to return back to the same
functional block (highlighted with blue arrows). Note that the size of
the dispatcher block exceeds 20 basic blocks while the functional
block consists of a single basic block.  

The oval nodes in
\autoref{fig:infloop} indicate basic blocks that are outside of the current
function. At basic block \texttt{0x41C79F},
function \texttt{ngx_time_sigsafe_update} is invoked. Due to the shortest path
heuristic, \sysname, tries to execute as few basic blocks as possible from
this function. In order to do so \sysname sets \texttt{ngx_time_lock} a non-zero value,
thus causing this function to return  quickly.
\sysname successfully synthesizes this payload in less than 5 minutes.

\subsection{Conditional statements}

\begin{figure}[t]
    \includegraphics[width=0.8\linewidth]{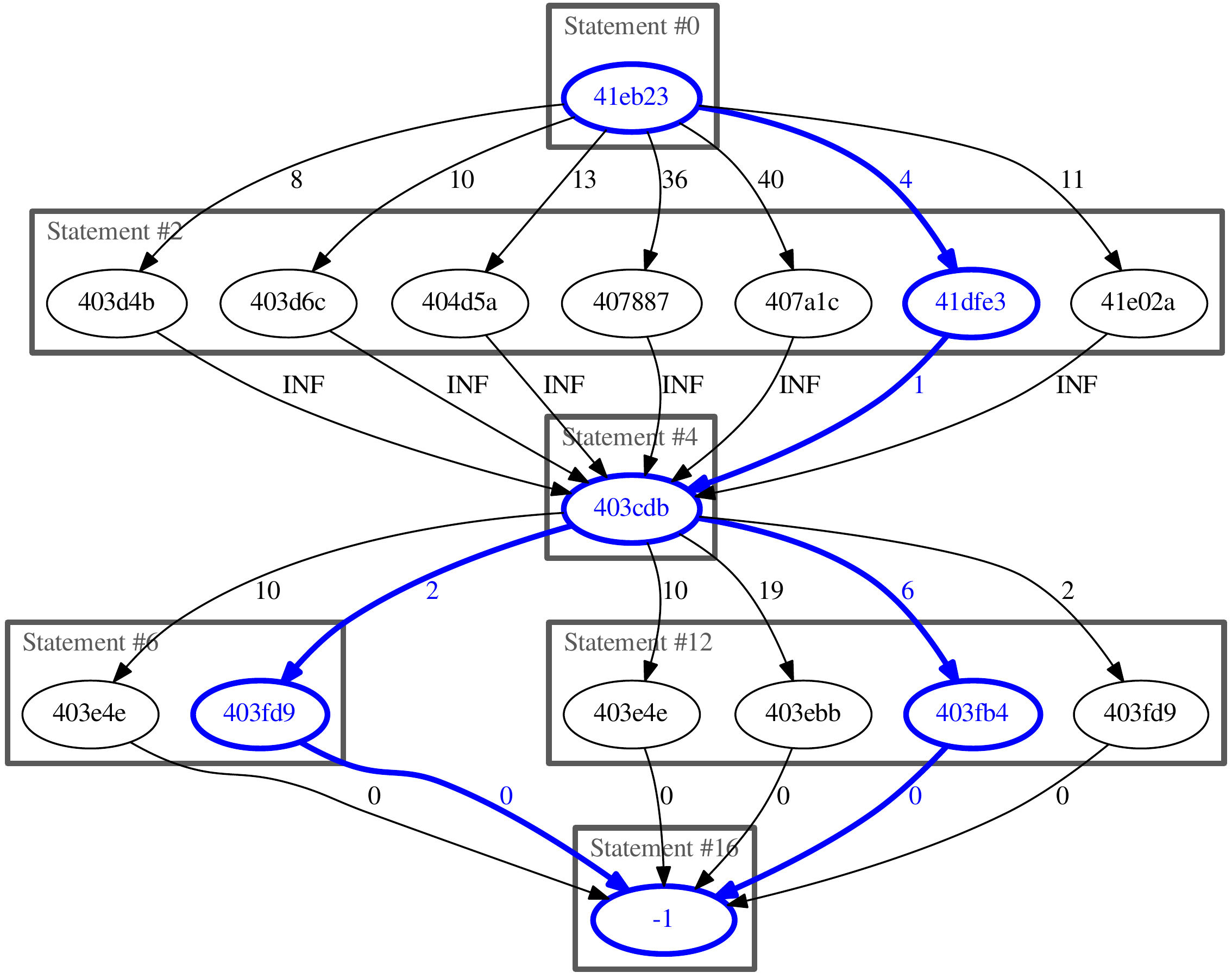}
\caption{A delta graph instance for an {\it ifelse} payload for nginx. The
first node is the {\it entry point}. Blue nodes and edges form 
the {\it minimum induced subgraph}, $H_k$. 
Statement $\#4$ is a conditional, execution branches into two 
statements.
Note that \sysname created this graph.
\vspace{-1em}
}
\label{fig:deltagraph}
\end{figure}

This case study shows an SPL if-else condition that implements a logical NOT.
That is, if register \texttt{\_\_r0} is zero, the payload sets \texttt{\_\_r1}
to one, otherwise \texttt{\_\_r1} becomes zero.
The execution trace starts at the beginning of 
\texttt{ngx_cache_manager_process_cycle}. This function is called through a
function pointer.
A part of the CFG starting from this function is shown in 
\aref{app:nginx-cfg}.
After trying $4$ mappings, \texttt{\_\_r0} and \texttt{\_\_r1} map to \texttt{rsi} and \texttt{r15} respectively. The resulting delta graph is
the shown in \autoref{fig:deltagraph}.

As we mentioned in \autoref{sec:simu}, when \sysname encounters a
functional block for a conditional statement, it \emph{clones} the current
state of the symbolic execution and the two clones independently continue the
execution. The constraints up to the conditional jump are the following:

\begin{minipage}[t]{\textwidth}
\centering
\begin{minted}[fontsize=\footnotesize, mathescape=true, framesep=1mm, baselinestretch=1.0,]{text}
    0x41eb23 : $rdi = ngx_cycle_t* cycle
    0x40f709 : *(ngx_event_flags + 1) == 0x2
    0x41dfe3 : __r0 = rsi = 0x0
    0x403cdb : $r15 = 0x1
               ngx_module_t ngx_core_module.index = 0
               $alloca_1 = *cycle
               ngx_core_conf_t* conf_ctx = 
                   *$alloca_1 + ngx_core_module.index * 8
    0x403d06 : test rsi, rsi (__r0 != 0)
    0x403d09 : jne  0x403d1b <ngx_set_environment+64>
\end{minted}
\vspace{1em}
\end{minipage}

If the condition is false and the jump is {\it not} taken, the following
constraints are also added to the state.

\begin{minipage}[t]{\textwidth}
\centering
\begin{minted}[fontsize=\footnotesize, framesep=1mm, baselinestretch=1.0,]{text}
    0x403d0b : conf_ctx->environment != 0
    0x403fd9 : __r1 = *($stack - 0x178) = 1; 
\end{minted}
\vspace{1em}
\end{minipage}

When the condition is true, the execution trace will follow the ``taken'' 
branch of the trace. In this case the shortest path to the next functional 
block is 
$403d1b \rightarrow 403d3d \rightarrow 403d4b \rightarrow 
 403d54 \rightarrow 403d5a \rightarrow 403fb4$ 
with a total length $6$. Unfortunately, this cannot be used as a dispatcher
block,
due to an exception that is raised at $403d4b$. The register \texttt{rsi}, is
$1$ and therefore when we attempt to execute the following instruction:
\texttt{cmp  BYTE PTR [rsi], 54h}, we essentially try to dereference address $1$.
\sysname is aware of this exception, so it discards the current path and tries
with the second shortest path.
The second shortest path has length $7$ and avoids the problematic block: 
$403d1b \rightarrow 403d8b \rightarrow 4050ba \rightarrow 40511c \rightarrow
 40513a \rightarrow 403d9c \rightarrow 403da5 \rightarrow403fb4$. This results
in a new set of constraints as shown below:

\begin{minipage}[t]{\textwidth}
\centering
\begin{minted}[fontsize=\footnotesize, mathescape=true, framesep=1mm, baselinestretch=1.0,]{text}
    0x403d1b : conf_ctx->env.elts = &elt (ngx_array_t*)
               conf_ctx->env.nelts == 0
    0x4050ba : conf_ctx->env.nelts != $alloca_2->env.nalloc
    0x40511c : conf_ctx->env.nelts += 1
    0x40513a : $ret = conf_ctx->env.elts +
                      conf_ctx->env.nelts*conf_ctx->env.size
    0x403d9c : $ret != 0
    0x403da5 : conf_ctx->env.nelts != 0
    0x403fb4 : __r1 = r15 = 0
\end{minted}
\vspace{1em}
\end{minipage}

\section{Discussion and Future Work} \label{sec:future}

Our prototype demonstrates the feasibility and scalability of automatic
construction of BOP chains through a high level language. However, we note
some potential optimizations that we will consider for future versions of
\sysname.

\sysname is limited by the \emph{granularity} of basic blocks.
That is, a combination of basic blocks could potentially lead to
the execution of a desired SPL statement, while individual blocks
might not.
Take for instance an instruction that sets a virtual register to $1$.
Assume that a basic block initializes \texttt{rcx} to $0$, while the
following block increments it by $1$; a pattern commonly encountered in loops.
Although there is no functional block that directly
sets \texttt{rcx} to $1$, the combination of the previous two has the
desired effect.
\sysname can be expanded to address this issue if the basic blocks are
coalesced into larger blocks that result in a new CFG.
\sysname sets several upper bounds defined by user inputs.
These configurable bounds include the upper limit of (i) SPL payload
permutations ($P$), (ii) length of continuous blocks ($L$), (iii) of minimum
induced subgraphs extracted from the delta graph ($N$), and (iv) dispatcher
paths between a pair of functional blocks ($K$).  These upper bounds along with
the timeout for symbolic execution, reduce the search space, but prune some
potentially valid solutions.  The evaluation of higher limits may result in
alternate or more solutions being found by \sysname.

\section{Conclusion} \label{sec:concl}

Despite the deployment of strong control-flow hijack defenses such as CFI or
shadow stacks, data-only code reuse attacks remain possible. So far, configuring
these attacks relies on complex manual analysis to satisfy restrictive
constraints for execution paths.

Our \sysname mechanism automates the analysis of the remaining attack surface
and synthesis of exploit payloads. To abstract complexity from target programs
and architectures, the payload is expressed in a high-level language. Our novel
code reuse technique, \emph{Block Oriented Programming}, maps statements of the
payload to functional basic blocks. Functional blocks are stitched together
through dispatcher blocks that satisfy the program CFG and avoid clobbering
functional blocks.  To find a solution for this NP-hard problem, we develop
heuristics to prune the search space and to evaluate the most probable paths
first.

The evaluation demonstrates that the majority of $13$ payloads, ranging from
typical exploit payloads to loops and conditionals are successfully mapped 81\%
of the time across $10$ programs.  Upon acceptance, we will release the source
code of our proof of concept prototype along with all of our evaluation results.
The prototype is available at \url{https://github.com/HexHive/BOPC}.

\section{Acknowledgments}

We thank the anonymous reviewers for their insightful comments. This
research was supported by ONR awards N00014-17-1-2513,
N00014-17-1-2498, by NSF CNS-1408880, CNS-1513783, CNS-1801534,
CNS-1801601, and a gift from Intel corporation. Any opinions,
findings, and conclusions or recommendations expressed in this
material are those of the authors and do not necessarily reflect the
views of our sponsors.

\bibliographystyle{ACM-Reference-Format}
\bibliography{references}
\appendix

\section{Extended Backus-Naur Form of SPL} \label{app:ebnf}

\setlength{\grammarparsep}{2pt plus 1pt minus 1pt} %
\setlength{\grammarindent}{6em} %
\begin{tcolorbox}
\begin{grammar}
<SPL>   ::= \textbf{void} \textbf{payload( )} \textbf{\{} <stmts> \textbf{\}}

<stmts> ::= $($<stmt> | <label>$)$* <return>?

<stmt>  ::= <varset> | <regset> | <regmod> | <call> \alt
                        <memwr>  | <memrd>  | <cond>   | <jump> \\
              
<varset> ::= \textbf{int64} <var> \textbf{=} <rvalue>\textbf{;}
       \alt \textbf{int64*} <var> \textbf{= \{}<rvalue> $($\textbf{,} 
       <rvalue>$)$*\textbf{\}}\textbf{;}
       \alt \textbf{string} <var> \textbf{=} <str>\textbf{;}
        
<regset> ::= <reg> \textbf{=} <rvalue>\textbf{;}

<regmod> ::= <reg> <op>\textbf{=} <number>\textbf{;}

<memwr>  ::= \textbf{*}<reg> \textbf{=} <reg>\textbf{;}

<memrd>  ::= <reg> \textbf{=} \textbf{*}<reg>\textbf{;}

<call>   ::= <var>~\textbf{(} $($ $\epsilon$ | <reg> (\textbf{,} <reg>$)$* 
                        \textbf{);}

<label>  ::= <var>\textbf{:}

<cond>   ::= \textbf{if (}<reg> <cmpop> <number>\textbf{)} 
             \textbf{goto} <var>\textbf{;}

<jump>   ::= \textbf{goto} <var>\textbf{;}

<return> ::= \textbf{returnto} <number>\textbf{;}
\\

<reg>    := `__r'<regid>

<regid>  := [0-7]

<var>    := [a-zA-Z_][a-zA-Z_0-9]*

<number> := (`+' | `-') [0-9]+ | `0x'[0-9a-fA-F]+

<rvalue> := <number> | `&' <var>

<str>    := [.]*

<op>     := `+' | `-' | `*' | `/' | `&' | `|' | `~' | `<<' | `<<'

<cmpop>  := `==' | `!=' | `>' | `>=' | `<' | `<='

\end{grammar}
\end{tcolorbox}

\section{Stitching BOP Gadgets is NP-Hard} \label{app:np-proof}

We present the NP-hardness proof for the BOP Gadget stitching problem. This 
problem reduces to the problem of finding the {\it minimum induced 
subgraph} $H_k$ in a delta graph.
Furthermore, we show that this problem cannot even be approximated.
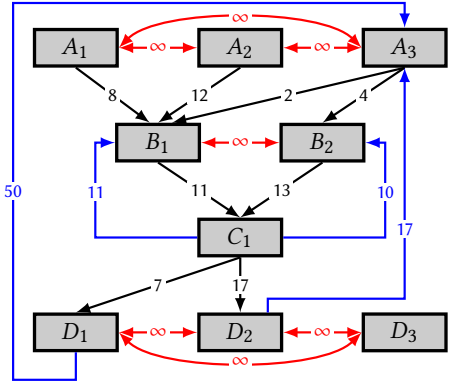
\begin{figure}[ht]
\begin{adjustbox}{width=.7\linewidth}
    \begin{tikzpicture}
        [
            unoptblock/.style={rectangle, fill=gray!40,draw,minimum width=30pt,minimum height=0em,very thick},
        myarrow/.style={->, >=latex, thick,color=black},
        mylabel/.style={midway,fill=white,inner sep=1pt},
            font=\sffamily
        ]
        \newcommand\connect[3]{\draw[->,draw] (#2) |- ++([yshift=5pt]$(#2)!0.5!(#3)$) node[above] {#1} -| (#3)}

        \node[unoptblock] (A21) {$A_1$};
        \node[unoptblock, right=of A21] (A22) {$A_2$};
        \node[unoptblock, right=of A22] (A23) {$A_3$};

        \node[unoptblock, below=20pt of $(A21.south)!0.5!(A22.south)$] (B1) {$B_1$};
        \node[unoptblock, below=20pt of $(A22.south)!0.5!(A23.south)$] (B2) {$B_2$};

        \node[unoptblock, below=20pt of $(B1.south)!0.5!(B2.south)$] (C1) {$C_1$};

        \node[unoptblock, below=20pt of $(C1.south)!0.5!(C1.south)$] (D2) {$D_2$};
        \node[unoptblock, left=of D2] (D1) {$D_1$};
        \node[unoptblock, right=of D2] (D3) {$D_3$};

        \draw[myarrow] (A21.south) -- ([xshift=-3pt]B1.north) node[mylabel] {\footnotesize 8};
        \draw[myarrow] (A22.south) -- (B1.north) node[mylabel] {\footnotesize 12};
        \draw[myarrow] (A23.south) -- (B2.north) node[mylabel] {\footnotesize 4};
        \draw[myarrow] (A23.south) -- ([xshift=5pt]B1.north) node[mylabel] {\footnotesize 2};
        \draw[myarrow] (B1.south) -- (C1.north) node[mylabel] {\footnotesize 11};
        \draw[myarrow] (B2.south) -- (C1.north) node[mylabel] {\footnotesize 13};
        \draw[myarrow] (C1.south) -- (D1.north) node[mylabel] {\footnotesize 7};
        \draw[myarrow] (C1.south) -- (D2.north) node[mylabel] {\footnotesize 17};

        \draw[myarrow,color=blue] (C1.west) -| ++(-1.3cm,0pt) |- (B1) node[mylabel,pos=0.24] {\footnotesize 11};
        \draw[myarrow,color=blue] (C1.east) -| ++(1.3cm,0pt) |- (B2) node[mylabel,pos=0.24] {\footnotesize 10};
        \draw[myarrow,color=blue] (D1.south) |- ++(-0.8cm,-10pt) |- ([yshift=9pt]A23.north) node[mylabel,pos=0.25] {\footnotesize 50} -| (A23.north);

        \draw[myarrow,color=blue] ([xshift=10pt]D2.north) |- ++(1.3cm,5pt)  -| (A23.south) node[mylabel, pos=0.65] {\footnotesize{17}};

        \draw[myarrow,<->,color=red] (A21) -- (A22) node[mylabel] {\footnotesize $\infty$};
        \draw[myarrow,<->,color=red] (A23) -- (A22) node[mylabel] {\footnotesize $\infty$};
        \draw[myarrow,<->,color=red] (B1) -- (B2) node[mylabel] {\footnotesize $\infty$};
        \draw[myarrow,<->,color=red] (D1) -- (D2) node[mylabel] {\footnotesize $\infty$};
        \draw[myarrow,<->,color=red] (D3) -- (D2) node[mylabel] {\footnotesize $\infty$};

        \draw[myarrow,<->,color=red, in=225,out=315,looseness=0.6] (D1.east) to (D3.west);
        \draw[myarrow,<->,color=red, in=135,out=45,looseness=0.6] (A21.east) to (A23.west);

        \node[above=0.5pt of A22,color=red,fill=white,inner sep=1pt] {\footnotesize$\infty$};
        \node[below=0.5pt of D2,color=red,fill=white,inner sep=1pt] {\footnotesize$\infty$};

    \end{tikzpicture}
\end{adjustbox}
\caption{
An delta graph instance. The nodes along the black edges form a
\emph{flat} delta graph. 
In this case, the {\it minimum induced subgraph}, $H_k$ 
is $A_3, B_1, C_1, D_1$, with a total
weight of $20$, which is also the {\it shortest path} from $A_3$ to $D_1$.
When delta graph is \emph{not} flat (assume that we add the blue edges),
the shortest path nodes constitute an induced subgraph with a total weight of 
$70$. However $H_k$ has total weight $34$ and contains $A_3, B_2, C_1, D_2$.
Finally, the problem of finding the minimum induced subgraph becomes
equivalent to finding a k-clique if we add the red edges with $\infty$ cost 
between all nodes in the same set.
} \label{fig:Hk}
\vspace{-1em}
\end{figure}

Let $\delta G$ be a {\it multipartite} directed weighted delta graph with 
$k$ sets. Our goal is to select \emph{exactly} one node (i.e., functional 
block) from each set and form the \emph{induced subgraph} $H_k$, such that 
the total weight of all of edges is \emph{minimized}:
\begin{equation}
        \min_{H_k \subset \delta G}\sum_{e \in H_k} distance(e)
\end{equation}

A $\delta G$ is \emph{flat}, when all edges from $i^{th}$ set are towards
$(i+1)^{th}$ set. The nodes and the black edges in \autoref{fig:Hk} are such
an example.
In this case, the minimum induced subgraph, is the minimum among all 
\emph{shortest paths} that start from some node in the first set and end 
in any node in the last set.
However, if the $\delta G$ is \emph{not} flat (i.e., the SPL payload contains 
jump statements, so edges from $i^{th}$ set can go anywhere), the shortest 
path approach does not work any more. Going back in \autoref{fig:Hk}, if we
make some loops (add the blue edges), the previous approach does not give the
correct solution.

It turns out that the problem is NP-hard if the $\delta G$ is not flat .
To prove this, we will use a reduction from {\it K-Clique}:
First we apply some equivalent transformations to the problem. Instead of 
having $K$ independent sets, we add an edge with $\infty$ weight between every 
pair on the same set, as shown in \autoref{fig:Hk} (red edges).
Then, the minimum weight K-induced subgraph $H_k$, cannot have two nodes from 
the same set, as this would imply that $H_k$ contains an edge with $\infty$ 
weight.

Let $R$ be an undirected un-weighted graph that we want to check whether it
has a $k$-clique. That is, we want to check whether $clique(R,k)$ is True or
not. Thus, we create a new directed graph $R\space'$ as follows:

\begin{itemize}
\item $R\space'$ contains all the nodes from $R$

\item $\forall$ edge $(u,v)\in R$, we add the edges $(u,v)$ and $(v,u)$ in
 $R\space'$ with $weight = 0$

\item $\forall$ edge $(u,v)\notin R$, we add the edges $(u,v)$ and $(v,u)$ in 
$R\space'$ with $weight = \infty$
\end{itemize}

Then we try to find the {\it minimum weight k-induced subgraph} $H_k$ in $R'$. 
It is true that:
$$\sum_{e \in H_k} weight(e) < \infty \Leftrightarrow clique(R,k) = True$$

$:\Rightarrow$
If the total edge weight of $H_k$ is not $\infty$, this implies that for every
pair of nodes in $H_k$, there is an edge with weight $1$ in $R\space'$ and thus
an edge in $R$. This by definition means that the nodes of  $H_k$ form a k-clique
in $R$.
Otherwise (the total edge weight of $H_k$ is $\infty$) it means that it does not
exist a set of $k$ nodes in $R\space'$ that has all edge weights $< \infty$.

$:\Leftarrow$
If $R$ has a k-clique, then there will be a set of $k$ nodes that are fully 
connected. This set of nodes will have no edge with $\infty$ weight in
$R\space'$. Thus, these nodes will form an induced subgraph of $R\space'$
and the total weight will be smaller than $\infty$.

This completes the proof that finding the minimum induced subgraph in 
$\delta G$ is NP-hard. However, no (multiplicative) approximation algorithm
does exists, as it would also solve the K-Clique problem (it must return 0 if
there is a K-Clique).

\section{SPL is Turing-complete} \label{app:spl-proof}
We present a constructive proof of Turing-completeness through
building an interpreter for Brainfuck~\cite{brainfuck}, a
Turing-complete language in the following listing.
This interpreter is written using SPL with a Brainfuck program provided
as input in the SPL payload.

\begin{minted}[fontsize=\footnotesize, mathescape=true,linenos,numbersep=0pt,frame=lines,framesep=2mm, baselinestretch=1.0,]{c}
  int64 *tape  = {0, 0, 0, 0, 0, 0, 0, 0, 0, 0};
  string input = ".+[.+]";
  __r0 = &tape;      // Data pointer
  __r2 = &input;     // Instruction pointer
  __r6 = 0;          // STDIN
  __r7 = 1;          // STDOUT
  __r8 = 1;          // Count arg for write/read
  NEXT:    __r1 = *__r2;
           if (__r1 != 0x3e) goto LESS;    // '>'
           __r0 += 1;
  LESS:    if (__r1 != 0x3c) goto PLUS;    // '<'
           __r0 -= 1;
  PLUS:    if (__r1 != 0x2b) goto MINUS;   // '+'
           *__r0 += 1;
  MINUS:   if (__r1 != 0x2d) goto DOT;     // '-'
           *__r0 -= 1;
  DOT:     if (__r1 != 0x2e) goto COMMA;   // '.'
           write(__r7, __r0, __r8);
  COMMA:   if (__r1 != 0x2c) goto OPEN;    // ','
           read(__r6, *__r0, __r8);
  OPEN:    if (__r1 != 0x5b) goto CLOSE;   // '['
           if (__r0 != 0) goto CLOSE;
           __r3 = 1;        // Loop depth counter
  FIND_C:  if (__r3 <= 0) goto CLOSE;
           __r2 += 1;
           __r1 = *__r2;
           if (__r1 != 0x5b) goto CHECK_C; // '['
           __r3 += 1;   
  CHECK_C: if (__r1 != 0x5d) goto FIND_C;  // ']'
           __r3 -= 1;
           goto FIND_C;
  CLOSE:   if (__r1 != 0x5d) goto END;     // ']'
           if (__r0 != 0) goto END;
           __r3 = 1;        // Loop depth counter
  FIND_O:  if (__r3 <= 0) goto END;
           __r2 -= 1;
           __r1 = *__r2;
           if (__r1 != 0x5b) goto CHECK_O; // '['
          __r3 -= 1;
  CHECK_O: if (__r1 != 0x5d) goto FIND_O;  // ']'
          __r3 += 1;
          goto FIND_O;
  END:    __r2 += 1;
  goto NEXT;
\end{minted}

\section{CFG of nginx after pruning} \label{app:nginx-cfg}
The following graph, is a portion of nginx's CFG that includes
function calls starting from
the function \texttt{ngx_cache_manager_process_cycle}.
The graph only displays functions which are up to 3 function calls deep to
simplify visualization.
Note the reduction in search space--which is a result of \sysname's pruning--as
this portion of the CFG
reduces to the small delta graph in \autoref{fig:deltagraph}.
\begin{figure}[h]
\includegraphics[width=.75\linewidth]{nginxcfg.pdf}
\label{fig:nginx-cfg}
\end{figure}

\end{document}